\documentstyle[12pt,epsfig,graphicx,color,amsmath,amsthm,amssymb]{article} 
\makeatletter \def\@cite#1#2{{[{#1}]\if@tempswa\typeout {IJCGA
warning: optional citation argument ignored: `#2'} \fi}}

\newcount\@tempcntc
\def\@citex[#1]#2{\if@filesw\immediate\write\@auxout{\string\citation{#2}}\fi
  \@tempcnta\z@\@tempcntb\m@ne\def\@citea{}\@cite{\@for\@citeb:=#2\do
    {\@ifundefined
       {b@\@citeb}{\@citeo\@tempcntb\m@ne\@citea\def\@citea{,}{\bf ?}\@warning
       {Citation `\@citeb' on page \thepage \space undefined}}%
    {\setbox\z@\hbox{\global\@tempcntc0\csname b@\@citeb\endcsname\relax}%
     \ifnum\@tempcntc=\z@ \@citeo\@tempcntb\m@ne
       \@citea\def\@citea{,}\hbox{\csname b@\@citeb\endcsname}%
     \else
      \advance\@tempcntb\@ne
      \ifnum\@tempcntb=\@tempcntc
      \else\advance\@tempcntb\m@ne\@citeo
      \@tempcnta\@tempcntc\@tempcntb\@tempcntc\fi\fi}}\@citeo}{#1}}
\def\@citeo{\ifnum\@tempcnta>\@tempcntb\else\@citea\def\@citea{,}%
  \ifnum\@tempcnta=\@tempcntb\the\@tempcnta\else
   {\advance\@tempcnta\@ne\ifnum\@tempcnta=\@tempcntb \else 
\def\@citea{--}\fi
    \advance\@tempcnta\m@ne\the\@tempcnta\@citea\the\@tempcntb}\fi\fi}
\makeatother

\def\boxit#1{\leavevmode\thinspace\hbox{\vrule\vtop{\vbox{\hrule%
        \vskip3pt\kern1pt\hbox{\vphantom{\bf/}\thinspace\thinspace%
        {\bf#1}\thinspace\thinspace}}\kern1pt\vskip3pt\hrule}\vrule}%
        \thinspace}
\def\Boxit#1{\noindent\vbox{\hrule\hbox{\vrule\kern3pt\vbox{
\advance\hsize-7pt\vskip-\parskip\kern3pt\bf#1 \hbox{\vrule height0pt
depth\dp\strutbox width0pt} \kern3pt}\kern3pt\vrule}\hrule}}

\newcommand{\gsim}{\lower.7ex\hbox{$\;\stackrel{\textstyle>}{\sim}\;$}}
\newcommand{\lsim}{\lower.7ex\hbox{$\;\stackrel{\textstyle<}{\sim}\;$}}

\allowdisplaybreaks[1]

\def\bad{\begin{aligned}[t]}
\def\ead{\end{aligned}}

\def\ifmath#1{\relax\ifmmode #1\else $#1$\fi}
\def\mgut{\ifmath{M_{\rm X}}}

\def\mLs{\ifmath{{{\bf m}_L^2}}}
\def\mes{\ifmath{{{\bf m}_e^2}}}

\def\Ae{\ifmath{{{\bf A}_e}}}

\begin{document}
\begin{titlepage}

\title{{\bf  Theoretical constraints on the rare tau
decays in the MSSM}}
\vskip3in \author{{\bf Alejandro Ibarra$^{1,2}$}, {\bf Tetsuo Shindou$^1$}
and {\bf Cristoforo Simonetto$^2$
\footnote{\baselineskip=16pt {\small E-mail addresses: {\tt
alejandro.ibarra@ph.tum.de, tetsuo.shindou@desy.de,\newline
cristoforo.simonetto@ph.tum.de}}}}
\hspace{3cm}\\
{\small $^1$ DESY,  Theory Group, Notkestrasse 85, D-22603 Hamburg, Germany}\\
{\small $^2$ Physik Department T30, Technische Universit\"at M\"unchen,}\\[-0.05cm]
{\it\normalsize James-Franck-Strasse, 85748 Garching, Germany}.
}  \date{}  \maketitle  \def\baselinestretch{1.15}
\begin{abstract}
\noindent 

The Minimal Supersymmetric Standard Model contains in general
sources of tau lepton flavour violation which 
induce the rare decays $\tau \rightarrow \mu \gamma$ and 
$\tau \rightarrow e \gamma$. We argue in this paper
that the observation of both rare processes would imply a lower
bound on the radiative muon decay of the form
${\rm BR}(\mu \rightarrow e \gamma)\gsim 
C\times {\rm BR}(\tau \rightarrow \mu \gamma)
{\rm BR}(\tau \rightarrow e \gamma)$. We estimate
the size of the constant $C$ without specifying 
the origin of the tau flavour violation in
the supersymmetric model and we discuss
the implications of our bound for future searches 
of rare lepton decays. In particular, we show that, for a wide class
of models, present $B$-factories
could discover either $\tau \rightarrow \mu \gamma$ or
$\tau \rightarrow e \gamma$, but not both. We also derive for
completeness the constant $C$ in the most general setup,
pursuing an effective theory approach.
\end{abstract}

\thispagestyle{empty}
\vspace*{0.2cm} \leftline{September 2008} \leftline{}

\vskip-16.5cm \rightline{DESY 08-114}
\rightline{TUM-HEP 697/08}

\end{titlepage}
\setcounter{footnote}{0} \setcounter{page}{1}
\newpage
\baselineskip=20pt

\noindent

\section{Introduction}

The existence of three generation of fermions with identical
gauge quantum numbers allows in principle electromagnetic 
transitions from a heavy generation into a light generation.
These transitions have been observed in the hadronic sector
(such as in the exclusive $B\rightarrow K^* \gamma$ decay~\cite{Ammar:1993sh})
but not in the leptonic sector. There exist in fact very
stringent bounds on the branching ratios of the lepton
flavour violating processes, that are summarized in Table \ref{tab:bounds}
together with the projected sensitivity of future experiments
to these decays.

\begin{table}[t]
\begin{center}
\begin{tabular}{|c|c|c|}
\hline 
& present bound & projected bound \\\hline
$BR(\mu\rightarrow e\gamma)$ & $1.2\times 10^{-11}$~\cite{Brooks:1999pu}&
$10^{-13}$~\cite{MEG} \\
$BR(\tau\rightarrow e\gamma)$ & $1.1\times 10^{-7}$~\cite{Aubert:2005wa}& 
$10^{-9}$~\cite{superB} \\
$BR(\tau\rightarrow \mu\gamma)$ & $4.5\times 10^{-8}$~\cite{Hayasaka:2007vc}& 
$10^{-9}$~\cite{superB} \\
\hline
\end{tabular}
\end{center}
\caption{\small Present and projected bounds on the rare lepton decays.}
\label{tab:bounds}
\end{table}

The puzzling difference between the hadronic sector
and the leptonic sector is very nicely explained in the framework 
of the Standard Model. The GIM mechanism~\cite{Glashow:1970gm} requires
that the decay rate for any flavour violating process 
is suppressed by the mass differences of the fermions
circulating in the loop over the $W$ boson mass. In 
the case of the leptonic transitions, the particles
circulating in the loop are neutrinos. Therefore, in the view 
of the tiny mass differences inferred from neutrino oscillation
experiments, the resulting decay rates are 
${\rm BR}(\tau\rightarrow \mu \gamma)\sim 10^{-54}$,
${\rm BR}(\mu\rightarrow e \gamma)\sim10^{-57}$, 
${\rm BR}(\tau\rightarrow e \gamma)\sim 10^{-57}$~\cite{meg_SM},
in agreement with the observations.

Nevertheless, the Standard Model is believed to be
an effective theory and new degrees of freedom are
expected to arise at some unspecified energy scale
between the electroweak scale and the Planck scale.  
Generically, the new degrees of freedom will couple
to the lepton doublets, potentially inducing new
sources of flavour violation. Therefore, the Standard Model
Lagrangian should be extended with higher-dimensional
effective operators to account for the flavour violation 
induced at low energies. The general expression for the 
electromagnetic transition amplitude $l_j \rightarrow l_i \gamma^*$ reads:
\begin{equation}
T=-e\, \epsilon^*_{\lambda} \bar u_i(p-q)
\left\{(f^{ji}_{E0}+\gamma_5f^{ji}_{M0})
\gamma_\nu (q^2 g^{\lambda\nu}-q^\lambda q^\nu)+ 
(f^{ji}_{M1}+\gamma_5f^{ji}_{E1})
i m_j \sigma^{\lambda \nu} q_\nu\right\} u_j(p)
\label{transition}
\end{equation}
where $p$ and $m_j$ are the momentum and the mass of the 
decaying lepton $l_j$, $q$ and $\epsilon^\lambda$ are the momentum 
and the polarization of the outgoing photon, and $f^{ji}_{E0}$, $f^{ji}_{M0}$,
$f^{ji}_{E1}$, $f^{ji}_{M1}$ are the different electromagnetic
form factors. If the photon is on shell, only the dipole 
operators contribute to the decay, which has a branching ratio
\begin{equation}
{\rm BR}(l_j\rightarrow l_i \gamma)=\frac{96 \pi^3 \alpha}{G_F^2}
(|f^{ji}_{E1}|^2+|f^{ji}_{M1}|^2) 
{\rm BR}(l_j\rightarrow l_i \nu_j \bar \nu_i)\,.
\label{BR-eff}
\end{equation}

The size and flavour structure of the form factors is completely
unknown. However, as we will show in this paper, there
exist correlations among the form factors that will eventually
translate into theoretical constraints on the branching ratios
of the rare processes. We will show that, barring cancellations, 
the following bound holds for any given model:
\begin{equation}
{\rm BR}(\mu \rightarrow e \gamma)\gsim 
C\times {\rm BR}(\tau \rightarrow \mu \gamma)
{\rm BR}(\tau \rightarrow e \gamma)\;,
\label{master}
\end{equation}
where the constant $C$ depends on the particular details of the model. 
As we will see, this bound has interesting implications for the 
searches for rare tau decays in present and future experiments.

Clearly, the more assumptions are imposed onto the model,
the more restrictive the bound becomes. In a previous paper 
\cite{Ibarra:2008uv} we derived the value of the constant 
$C$ for the case of the  supersymmetric see-saw model and 
we reached the interesting conclusion 
that, for large regions of the mSUGRA parameter space, present
$B$-factories could either discover $\tau \rightarrow \mu \gamma$
or $\tau \rightarrow e \gamma$ but not both. In the present work 
we extend this analysis to more general models. In Section 2 
we will show our analysis for the Minimal Supersymmetric Standard
Model (with R-parity conserved) and in Section 3 for a general effective 
theory described by Eq.~(\ref{transition}). Finally, in
Section 4 we will present our conclusions. 

\section{Minimal Supersymmetric Standard Model}

The scalar sector of the Minimal Supersymmetric Standard Model (MSSM)
contains additional sources of lepton flavour violation 
in the soft supersymmetry (SUSY) breaking Lagrangian~\cite{offdiag}, 
which reads
\begin{eqnarray}  
-{\cal L}^{\rm lep}_{\rm soft}&=&\ 
(\mLs)_{ij} \widetilde L^*_i  \widetilde L_j\ +
(\mes)_{ij} \widetilde e^*_{Ri}  \widetilde e_{Rj}\ + 
 \left(\Ae_{ij} \widetilde e^*_{Ri} H_d \widetilde L_j +
{\rm h.c.}\right).
\end{eqnarray} 
In this Lagrangian
$\widetilde L_i$ and $\widetilde e_{Ri}$ 
are the supersymmetric partners of the left-handed lepton doublets and
right-handed charged leptons, respectively,
$\mLs$ and $\mes$  are their corresponding soft mass matrices squared,
and $\Ae$ is the charged lepton soft trilinear term.

After the electroweak symmetry breaking, left-handed and right-handed
charged sleptons mix. The corresponding $6\times 6$ mass
matrix can be parametrized as
\begin{equation}
{\cal M}_{\tilde{e}}^2=
\begin{pmatrix}
m_{L}^2 \,{\mathbb I}_3+\Delta^{\rm (LL)}&
m_{LR}\, m^{\rm lep}
+\Delta^{\rm (LR)}\\
m_{RL}\, m^{\rm lep}
+\Delta^{\rm (RL)}
&m_{R}^2\,{\mathbb I}_3+\Delta^{\rm (RR)}
\end{pmatrix} \;,
\label{charged-sleptons}
\end{equation}
where $m_L$ and $m_R$ are the average masses of the left and right-handed
charged sleptons, respectively, $m^{\rm lep}={\rm diag}(m_e,m_\mu,m_\tau)$
is the charged lepton mass matrix, and $m_{LR}=m^*_{RL}$ is the
average left-right mixing term. It approximately reads
$m_{LR}\sim {\widetilde m}\, \tan\beta$, being 
${\widetilde m}$ a SUSY mass scale.
On the other hand, in the absence of right-handed neutrino superfields,
the sneutrino mass matrix is just a $3\times 3$ matrix that can be 
parametrized in an analogous way:
\begin{equation}
{\cal M}_{\tilde{\nu}}^2=
\bar{m}_{L}^2\,{\mathbb I}_3+\Delta^{\rm (LL)}\;,
\label{sneutrinos}
\end{equation}
with $\bar{m}_{L}$ the average sneutrino mass.
With these definitions, the $3\times 3$ matrices $\Delta^{\rm (LL)}$, 
$\Delta^{\rm (RR)}$, $\Delta^{\rm (LR)}$ and 
$\Delta^{\rm (RL)}$  encode all the
flavour structure of the soft SUSY breaking terms.

The branching ratios for the different radiative decays can be 
straightforwardly computed from the general formulas existing 
in the literature~\cite{Hisano:1995cp}. Nevertheless, in order to understand 
qualitatively the results, it is useful to derive approximate 
expressions for the cumbersome formulas of the branching ratios.
We will use, however, the exact expressions for our numerical analysis.

We will adopt in this paper the mass insertion approximation,
which consists on treating the small off-diagonal elements of the 
soft terms as insertions in the sfermion propagators in the 
loops~\cite{mass-insertion}.
Then, the branching ratio for the radiative lepton decays can be 
schematically written as:
\begin{equation}
{\rm BR}(l_j\rightarrow l_i \gamma)= |f_{ij}^{(1)} \Delta_{ij} +
f_{ij}^{(2)} \Delta_{ik}\Delta_{jk}^*+...|^2,~~~~~ k\neq i,j\;,
\end{equation}
where $\Delta_{ij}$ denotes generically any mass insertion.
In this perturbative expansion, the first term corresponds to 
the single mass insertion, the
second, to the double mass insertion, etc. It is apparent from this
expression that, barring unnatural cancellations, 
the observation of two radiative rare decays
implies a non-vanishing rate for the third one. For instance,
the observation of  $\tau\rightarrow \mu \gamma$ 
and $\tau\rightarrow e \gamma$ would imply, barring cancellations, 
a lower bound on the rate of $\mu\rightarrow e\gamma$:
\begin{equation}
{\rm BR}(\mu\rightarrow e\gamma)\gsim 
\frac{|f_{e \mu}^{(2)}|^2}{|f_{\mu\tau}^{(1)}|^2|f_{e \tau}^{(1)}|^2}
{\rm BR}(\tau\rightarrow \mu\gamma){\rm BR}(\tau\rightarrow e\gamma)\;,
\label{bound-fs}
\end{equation}
which is saturated when $\Delta_{e \mu}=0$, {\it i.e.} when the decay 
rate is dominated by the double mass insertion. This equation is the
supersymmetric realization of the general bound Eq.~(\ref{master}).
The reason for this
correlation among the rare tau decays can be traced back to
the fact that the observation of both rare tau decays would
imply that all family lepton numbers are violated in
nature, and thus there is no symmetry reason forbidding
the process $\mu\rightarrow e\gamma$. Although this rationale
can be applied to any two rare processes to infer a lower bound on 
the rate of the third one, in the view of the stringent present
constraint on $\mu\rightarrow e\gamma$ and the excellent prospects
to improve the experimental sensitivity to this process in the
near future, we will just discuss in detail the correlation 
Eq.~(\ref{bound-fs}) 
for $\mu\rightarrow e\gamma$ and the implications of this
bound for future searches of rare tau decays.

Assuming that one source of flavour violation, LL, RR, RL or LR,
dominates, the branching ratios for the rare tau decays can
be written, respectively, as 
\begin{align}
&{\rm BR}(\tau\rightarrow \mu \gamma)
\sim \frac{\alpha^3}{G_F^2}
\left|\frac{\Delta^{\rm (LL)}_{\mu\tau}}{{\widetilde m}_{\rm (LL)}^4},
\frac{\Delta^{\rm (RR)}_{\mu\tau}}{{\widetilde m}_{\rm (RR)}^4},
\frac{\Delta^{\rm (RL)}_{\mu\tau}}{{\widetilde m}_{\rm (RL)}^3 m_\tau\tan\beta},
\frac{\Delta^{\rm (LR)}_{\mu\tau}}{{\widetilde m}_{\rm (LR)}^3 m_\tau\tan\beta}
\right|^2\tan^2\beta 
\;{\rm BR}(\tau \rightarrow \mu \nu_\tau \bar \nu_\mu) \;,\nonumber\\
&{\rm BR}(\tau\rightarrow e \gamma)
\sim \frac{\alpha^3}{G_F^2}
\left|\frac{\Delta^{\rm (LL)}_{e \tau}}{{\widetilde m}_{\rm (LL)}^4},
\frac{\Delta^{\rm (RR)}_{e \tau}}{{\widetilde m}_{\rm (RR)}^4},
\frac{\Delta^{\rm (RL)}_{e \tau}}{{\widetilde m}_{\rm (RL)}^3 m_\tau \tan\beta},
\frac{\Delta^{\rm (LR)}_{e \tau}}{{\widetilde m}_{\rm (LR)}^3 m_\tau \tan\beta}
\right|^2\tan^2\beta 
\;{\rm BR}(\tau \rightarrow e \nu_\tau \bar \nu_e) \;,
\label{1MI}
\end{align}
where ${\rm BR}(\tau\rightarrow \mu \nu_\tau \bar \nu_\mu)\simeq 0.17$,
${\rm BR}(\tau\rightarrow e \nu_\tau \bar \nu_e)\simeq 0.18$ and
${\widetilde m}_{\rm (LL)}$, ${\widetilde m}_{\rm (RR)}$,
${\widetilde m}_{\rm (RL)}$ and ${\widetilde m}_{\rm (LR)}$
are mass scales of the order of typical SUSY masses.

Since ${\rm BR}(\tau\rightarrow \mu  \gamma)$ and 
${\rm BR}(\tau\rightarrow e \gamma)$ can be, each of them,
generated by four different mass insertions, there
are 16 possible combinations for the double mass
insertion that induces $\mu\rightarrow e \gamma$.
The lower bound on the rate for $\mu\rightarrow e\gamma$ is
approximately given by
\begin{equation}
{\rm BR}(\mu\rightarrow e \gamma)\gsim \frac{\alpha^3}{G_F^2}
\left|\frac{\Delta^{\rm (X)}_{e \tau}\Delta^{\rm (Y)*}_{\mu\tau}}
{\widetilde m_{\rm (X,Y)}^6} h^{\rm (X,Y)}\right|^2 \tan^2\beta\;,
\label{2MI}
\end{equation}
where X, Y=LL, RR, LR, RL and ${\widetilde m}_{\rm (X,Y)}$ 
is another mass scale of
the order of typical SUSY masses, in general different from 
${\widetilde m}_{\rm (LL)},{\widetilde m}_{\rm (RR)}, 
{\widetilde m}_{\rm (LR)}, {\widetilde m}_{\rm (RL)}$.
On the other hand, $h^{\rm (X,Y)}$ is a factor that depends crucially
on which are the particular mass insertions considered and that is listed
in Table~\ref{tab:doubleMI} for all the 16 combinations.
It takes non-trivial values in those combinations
that require a left-right mass insertion in the stau
propagator, thus introducing a factor 
$(m_\tau \tan\beta)/{\widetilde m_{\rm (X,Y)}}$, and those combinations
where the chirality flip occurs in the gaugino propagator,
introducing a factor ${\widetilde m_{\rm (X,Y)}}/(m_\mu \tan\beta)$.

\begin{table}
\begin{center}
\begin{tabular}{|c|cccc|}
\hline 
& LL & RR & LR & RL \\ \hline
LL
& 1 
&$\frac{m_\tau}{m_\mu}$
&$\frac{m_\tau \tan\beta}{\widetilde m_{\rm (LL,LR)}}$
&$\frac{\widetilde m_{\rm (LL,RL)}}{m_\mu \tan\beta}$  \\
RR
& $\frac{m_\tau}{m_\mu}$ 
& 1
& $\frac{\widetilde m_{\rm (RR,LR)}}{m_\mu \tan\beta}$ 
& $\frac{m_\tau \tan\beta}{\widetilde m_{\rm (RR,RL)}}$\\
LR
& $\frac{m_\tau \tan\beta}{\widetilde m_{\rm (LR,LL)}}$
& $\frac{\widetilde m_{\rm (LR,RR)}}{m_\mu \tan\beta}$  
& 1
& $\frac{m_\tau}{m_\mu}$\\
RL
& $\frac{\widetilde m_{\rm (RL,LL)}}{m_\mu \tan\beta}$  
& $\frac{m_\tau \tan\beta}{\widetilde m_{\rm (RL,RR)}}$
& $\frac{m_\tau}{m_\mu}$
& 1\\
\hline
\end{tabular}
\end{center}
\caption{Values of the factor $h^{\rm (X,Y)}$, defined in Eq.~(\ref{2MI}),
for all the 16 possible combinations inducing the process $\mu\to e\gamma$ 
through a double mass insertion diagram.}
\label{tab:doubleMI}
\end{table}

Using Eqs.(\ref{1MI},\ref{2MI}) it is straightforward to derive bounds
of the form ${\rm BR}(\mu \rightarrow e \gamma)\gsim 
C\times {\rm BR}(\tau \rightarrow \mu \gamma)
{\rm BR}(\tau \rightarrow e \gamma)$
for all the 16 possibilities. We can classify the results 
in four classes, each of them having the same dependence
on $\tan\beta$, the fermion masses and the overall size
of the scalar masses, which are the three parameters to
which the constant $C$ is most sensitive to:
\begin{itemize}
\item{Class I: 
$\Delta_{e \tau}^{\rm (LL)}\Delta_{\mu \tau}^{\rm (LL)*}$ and
$\Delta_{e \tau}^{\rm (RR)}\Delta_{\mu \tau}^{\rm (RR)*}$.
\begin{equation}
{\rm BR}(\mu\rightarrow e\gamma) \gsim \frac{G_F^2}{\alpha^3 \tan^2\beta}
\left[
\frac{m^{8}_{\rm (LL)} m^{8}_{\rm (LL)}}{m^{12}_{\rm (LL,LL)}},
\frac{m^{8}_{\rm (RR)} m^{8}_{\rm (RR)}}{m^{12}_{\rm (RR,RR)}}
\right]
\frac{{\rm BR}(\tau\rightarrow \mu\gamma)}
{{\rm BR}(\tau\rightarrow \mu \nu_\tau \bar \nu_\mu)}
\frac{{\rm BR}(\tau\rightarrow e\gamma)}
{{\rm BR}(\tau\rightarrow e \nu_\tau \bar \nu_e)} \;.
\label{class1}
\end{equation}}
\item{Class II: 
$\Delta_{e \tau}^{\rm (LL)}\Delta_{\mu \tau}^{\rm (RR)*}$,
$\Delta_{e \tau}^{\rm (RR)}\Delta_{\mu \tau}^{\rm (LL)*}$,
$\Delta_{e \tau}^{\rm (LR)}\Delta_{\mu \tau}^{\rm (RR)*}$,
$\Delta_{e \tau}^{\rm (RR)}\Delta_{\mu \tau}^{\rm (LR)*}$,
$\Delta_{e \tau}^{\rm (RL)}\Delta_{\mu \tau}^{\rm (LL)*}$ and
$\Delta_{e \tau}^{\rm (LL)}\Delta_{\mu \tau}^{\rm (RL)*}$.
\begin{eqnarray}
{\rm BR}(\mu\rightarrow e\gamma) &\gsim& \frac{G_F^2}{\alpha^3 \tan^2\beta}
\frac{m^2_\tau}{m^2_\mu}
\left[
\frac{m^{8}_{\rm (LL)} m^{8}_{\rm (RR)}}{m^{12}_{\rm (LL,RR)}},
\frac{m^{6}_{\rm (LR)} m^{8}_{\rm (RR)}}{m^{10}_{\rm (LR,RR)}},
\frac{m^{6}_{\rm (RL)} m^{8}_{\rm (LL)}}{m^{10}_{\rm (RL,LL)}}
\right] \nonumber \\
&&\frac{{\rm BR}(\tau\rightarrow \mu\gamma)}
{{\rm BR}(\tau\rightarrow \mu \nu_\tau \bar \nu_\mu)}
\frac{{\rm BR}(\tau\rightarrow e\gamma)}
{{\rm BR}(\tau\rightarrow e \nu_\tau \bar \nu_e)}\;;
\label{class2}
\end{eqnarray}}
\item{Class III: 
$\Delta_{e \tau}^{\rm (LL)}\Delta_{\mu \tau}^{\rm (LR)*}$,
$\Delta_{e \tau}^{\rm (LR)}\Delta_{\mu \tau}^{\rm (LL)*}$,
$\Delta_{e \tau}^{\rm (RR)}\Delta_{\mu \tau}^{\rm (RL)*}$,
$\Delta_{e \tau}^{\rm (RL)}\Delta_{\mu \tau}^{\rm (RR)*}$,
$\Delta_{e \tau}^{\rm (LR)}\Delta_{\mu \tau}^{\rm (LR)*}$ and
$\Delta_{e \tau}^{\rm (RL)}\Delta_{\mu \tau}^{\rm (RL)*}$.
\begin{eqnarray}
{\rm BR}(\mu\rightarrow e\gamma) &\gsim& \frac{G_F^2}{\alpha^3}
m^4_\tau \tan^2\beta
\left[
\frac{m^{8}_{\rm (LL)} m^{6}_{\rm (LR)}}{m^{14}_{\rm (LR,LL)}},
\frac{m^{8}_{\rm (RR)} m^{6}_{\rm (RL)}}{m^{14}_{\rm (LR,LL)}},
\frac{m^{12}_{\rm (LR)}}{m^{12}_{\rm (LR,LR)}},
\frac{m^{12}_{\rm (RL)}}{m^{12}_{\rm (RL,RL)}}
\right]
\nonumber \\
&&
\frac{{\rm BR}(\tau\rightarrow \mu\gamma)} 
{{\rm BR}(\tau\rightarrow \mu \nu_\tau \bar \nu_\mu)}
\frac{{\rm BR}(\tau\rightarrow e\gamma)}
{{\rm BR}(\tau\rightarrow e \nu_\tau \bar \nu_e)} 
 \;.
\label{class3}
\end{eqnarray}}
\item{Class IV:
$\Delta_{e \tau}^{\rm (LR)}\Delta_{\mu \tau}^{\rm (RL)*}$ and
$\Delta_{e \tau}^{\rm (RL)}\Delta_{\mu \tau}^{\rm (LR)*}$.
\begin{equation}
{\rm BR}(\mu\rightarrow e\gamma) \gsim \frac{G_F^2}{\alpha^3}
\frac{m^6_\tau \tan^2\beta}{m^2_\mu}
\frac{m^{6}_{\rm (LR)} m^{6}_{\rm (RL)}}{m^{12}_{\rm (LR,RL)}}
\frac{{\rm BR}(\tau\rightarrow \mu\gamma)}
{{\rm BR}(\tau\rightarrow \mu \nu_\tau \bar \nu_\mu)}
\frac{{\rm BR}(\tau\rightarrow e\gamma)}
{{\rm BR}(\tau\rightarrow e \nu_\tau \bar \nu_e)} \;.
\label{class4}
\end{equation}}
\end{itemize}

The numerical values of the mass scales $m_{\rm (X)}$ and $m_{\rm (X,Y)}$
depend on the particular supersymmetric scenario considered.
Before presenting exact results for concrete SUSY benchmark points,
let us first derive rough numerical estimates of the bounds 
Eqs.(\ref{class1}-\ref{class4}).
To this end, we will make the approximation 
$m_{\rm (X)}= m_{\rm (X,Y)}={\widetilde m}$ for all X, Y.
Then, the previous
bounds read:
\begin{itemize}
\item{Class I:
\begin{equation}
{\rm BR}(\mu\rightarrow e\gamma) \gsim  9\times 10^{-10}
\left(\frac{\widetilde m}{200\,{\rm GeV}}\right)^4
\left(\frac{\tan\beta}{10}\right)^{-2}
\left(\frac{{\rm BR}(\tau\rightarrow \mu\gamma)}{4.5\times10^{-8}}\right)
\left(\frac{{\rm BR}(\tau\rightarrow e\gamma)}{1.1\times10^{-7}}\right)\;.
\label{class1-rough}
\end{equation}}
\item{Class II:
\begin{equation}
{\rm BR}(\mu\rightarrow e\gamma) \gsim  3\times 10^{-7}
\left(\frac{\widetilde m}{200\,{\rm GeV}}\right)^4
\left(\frac{\tan\beta}{10}\right)^{-2}
\left(\frac{{\rm BR}(\tau\rightarrow \mu\gamma)}{4.5\times10^{-8}}\right)
\left(\frac{{\rm BR}(\tau\rightarrow e\gamma)}{1.1\times10^{-7}}\right)\;.
\label{class2-rough}
\end{equation}}
\item{Class III
\begin{equation}
{\rm BR}(\mu\rightarrow e\gamma) \gsim  5 \times 10^{-14}
\left(\frac{\tan\beta}{10}\right)^{2}
\left(\frac{{\rm BR}(\tau\rightarrow \mu\gamma)}{4.5\times10^{-8}}\right)
\left(\frac{{\rm BR}(\tau\rightarrow e\gamma)}{1.1\times10^{-7}}\right)\;.
\label{class3-rough}
\end{equation}}
\item{Class IV
\begin{equation}
{\rm BR}(\mu\rightarrow e\gamma) \gsim 2\times 10^{-11}
\left(\frac{\tan\beta}{10}\right)^{2}
\left(\frac{{\rm BR}(\tau\rightarrow \mu\gamma)}{4.5\times10^{-8}}\right)
\left(\frac{{\rm BR}(\tau\rightarrow e\gamma)}{1.1\times10^{-7}}\right)\;.
\label{class4-rough}
\end{equation}}
\end{itemize}
Notice that as $\tan\beta$ increases the bound becomes stronger for
Classes III and IV, while it becomes weaker for Classes I and II.
Notice also that for Classes III and IV the bound is not very sensitive
to the size of the SUSY masses, while for Classes I and II it becomes
stronger as the SUSY mass scale increases\footnote{One loop QED
corrections to the electric and magnetic dipole operators reduce
the theoretical prediction for ${\rm BR}(l_j\rightarrow l_i \gamma)$ 
by a factor $\left(1-\frac{8\alpha}{\pi}
\log\frac{\widetilde m}{m_j}\right)$~\cite{Czarnecki:2001vf}.
This correction makes the bounds Eqs.~(\ref{class1-rough}-\ref{class4-rough}) 
a 2-6\% stronger for $\widetilde m=100-1000$ GeV.}.

From these bounds it follows that if the rates for both rare tau 
decays were just below the present experimental bound, only 
the scenarios falling in Class III (and marginally in Class IV)
would be allowed. In contrast, for Class I the rate for 
$\mu\rightarrow e \gamma$ induced by the double mass insertion
would be much larger than the MEGA bound, unless $\tan\beta$ is very large and 
the soft masses are small (for $\tan\beta=50$, ${\widetilde m}$
has to be smaller than 150 GeV in order to satisfy the 
bound ${\rm BR}(\mu\rightarrow e\gamma)\leq 1.2\times 10^{-11}$ from MEGA).
On the other hand, scenarios falling in Class II would be excluded unless
a strong cancellation is taking place among the different contributions.
The same conclusion holds if both rare decays were accessible
to present $B$-factories, which requires ${\rm BR}(\tau\rightarrow l_i
\gamma)\gsim 10^{-8}$. Therefore, if both rare tau decays
were observed in present $B$-factories, the possible sources
of flavour violation would be restricted to Classes III and IV 
in most of the SUSY parameter space. Clearly, these conclusions
will become stronger if the MEG experiment at PSI reaches the
projected sensitivity of $10^{-13}$ on ${\rm BR}(\mu\rightarrow e \gamma)$
without finding a positive signal.

When interpreting the previous bounds one should bear in mind
that Eqs.(\ref{class1}-\ref{class4}) are 
proportional to very large powers of the masses. Therefore, the
numerical values of the bounds Eqs.(\ref{class1-rough}-\ref{class4-rough})
may vary one or two orders of magnitude even
if $m_{\rm (X)}\sim m_{\rm (X,Y)}$~\footnote{For instance,
if $m_{\rm (X)}=m_{\rm (Y)}=\sqrt{2} m_{\rm (X,Y)}$
 the bounds get relaxed by a factor 64, and conversely, if 
 $m_{\rm (X)}=m_{\rm (Y)}=1/\sqrt{2} m_{\rm (X,Y)}$ 
the bounds get strengthened by a factor 64.}. 
Nevertheless, given that the numerical
value of the result is typically more than two orders of magnitude
above or below the experimental bound this uncertainty usually does 
not alter our conclusions, except perhaps for Class IV. 
To check our general expectations we have analyzed in detail 
the SPS1a and SPS1b benchmark points
~\cite{Allanach:2002nj}, which correspond to ``typical''
mSUGRA points with intermediate and relatively high
values of $\tan\beta$, respectively. They are characterized
by five parameters at the Grand Unified Scale, 
$\mgut=2\times 10^{16}$ GeV, namely the
universal scalar mass ($m_0$), gaugino mass ($M_{1/2}$) and
trilinear term ($A_0$), $\tan\beta$ and the sign of $\mu$.
For the SPS1a (SPS1b) point, these parameters are
$m_0=100 (200)$ GeV, $M_{1/2}=250 (400)$ GeV, $A_0=-100 (0)$ GeV,
$\tan\beta=10 (30)$ and ${\rm sign}\;\mu=+$.

\begin{figure}
\begin{center}
\begin{tabular}{c}
\epsfig{figure=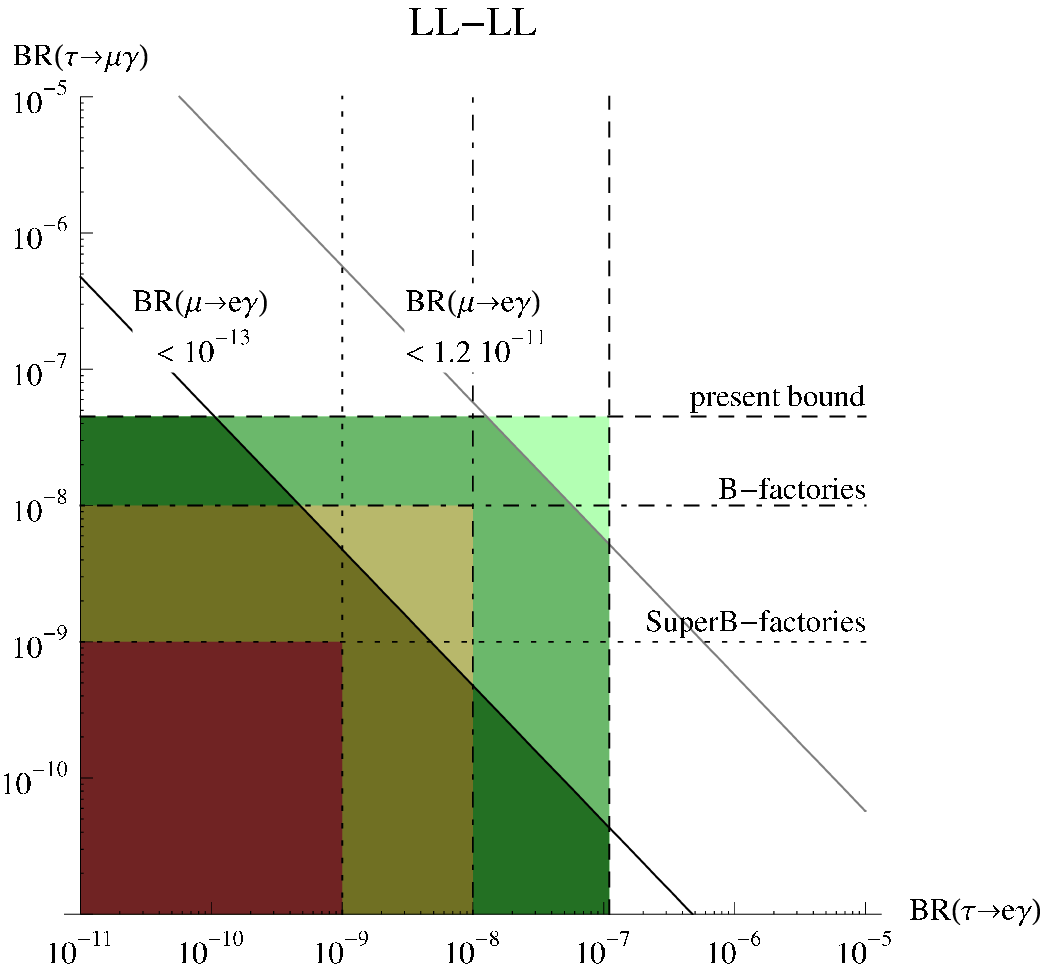,width=65mm} 
\epsfig{figure=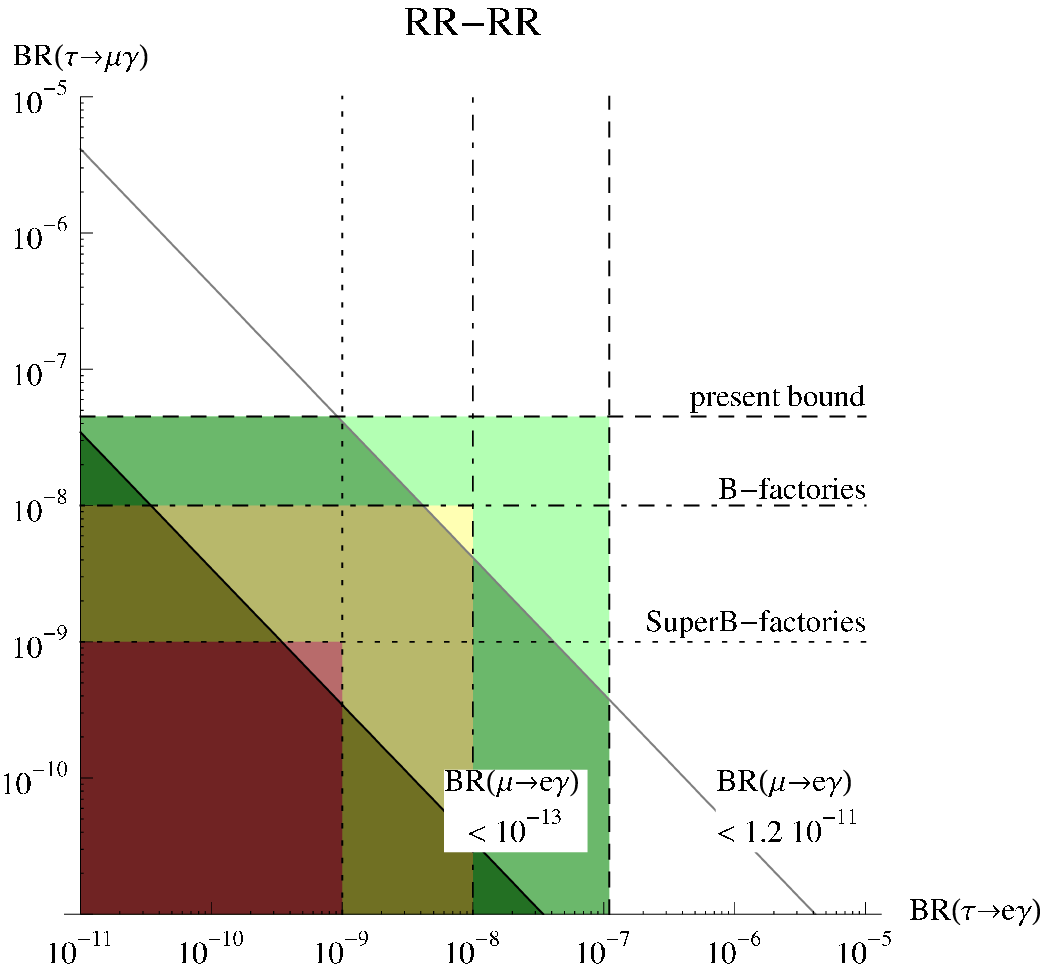,width=65mm}
\end{tabular}
\end{center}
\caption
{\small  Allowed values for the branching ratios
of the rare tau decays $\tau\rightarrow e\gamma$ and
$\tau\rightarrow \mu\gamma$ from present experiments
and from the bound ${\rm BR}(\mu\rightarrow e\gamma)\gsim
C \times {\rm BR}(\tau\rightarrow \mu\gamma){\rm BR}(\tau\rightarrow e\gamma)$
for the mass insertions falling in Class I (see text).
The area in green indicates the values of the branching
ratios that are accessible to present $B$-factories,
and in yellow, the ones accessible to future super$B$-factories.
Excluded regions are shown with light shading, whereas
allowed regions are shown with dark shading. 
The supersymmetric benchmark point SPS1a has been assumed.
}
\label{fig:classI}
\end{figure}

\begin{figure}
\begin{center}
\begin{tabular}{c}
\epsfig{figure=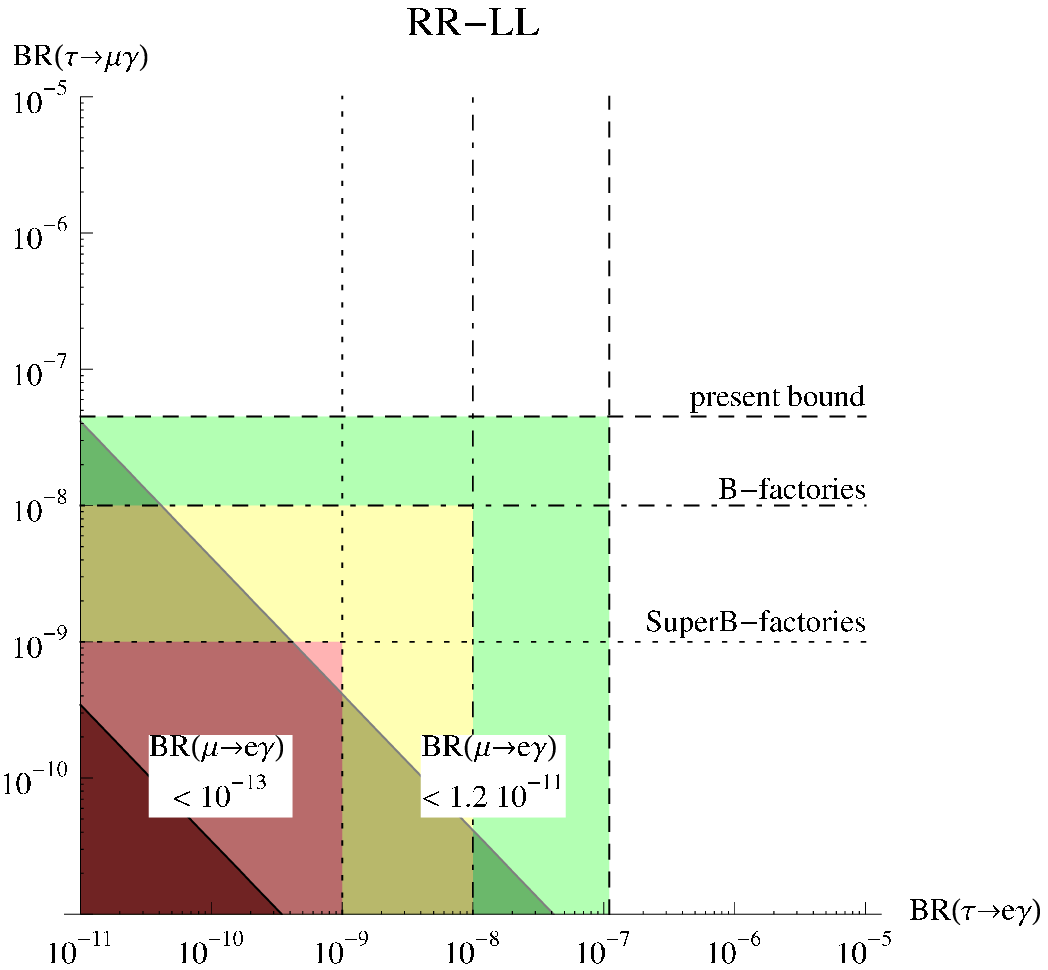,width=65mm} 
\epsfig{figure=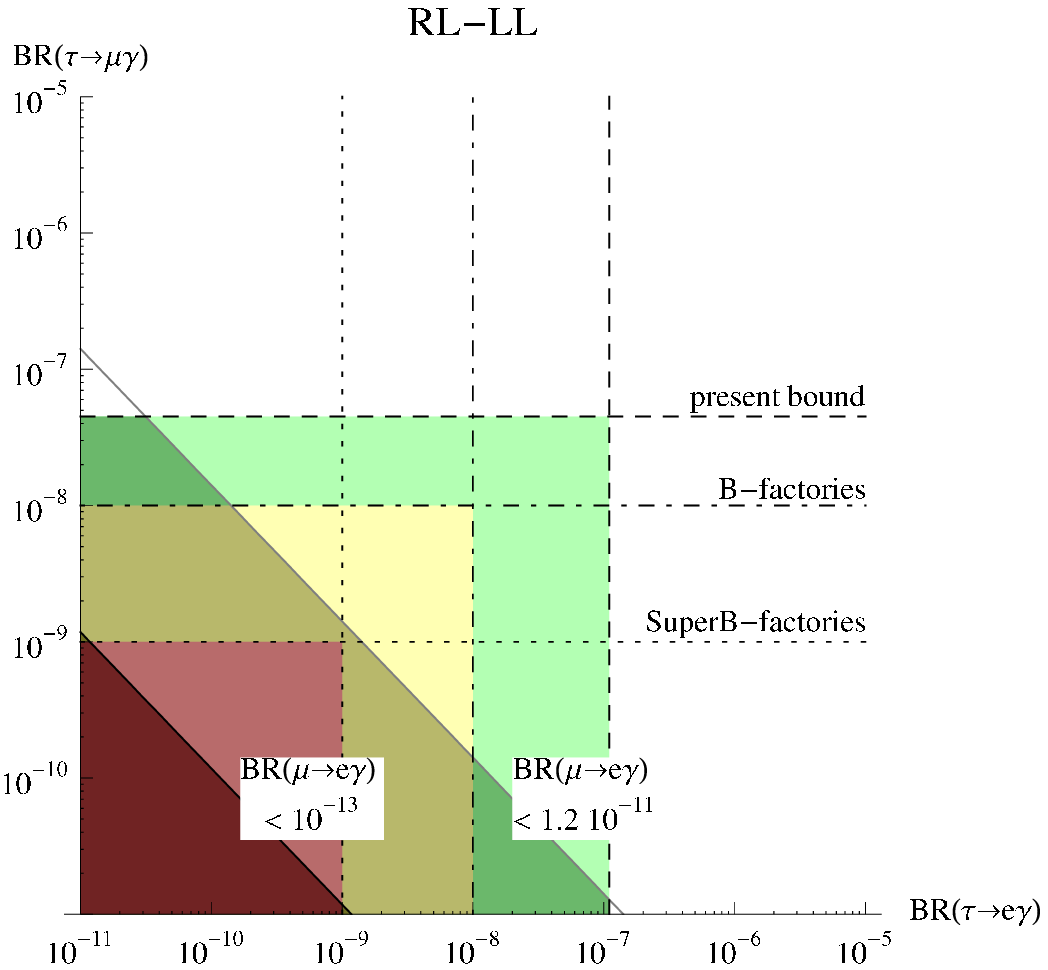,width=65mm}\\\\
\epsfig{figure=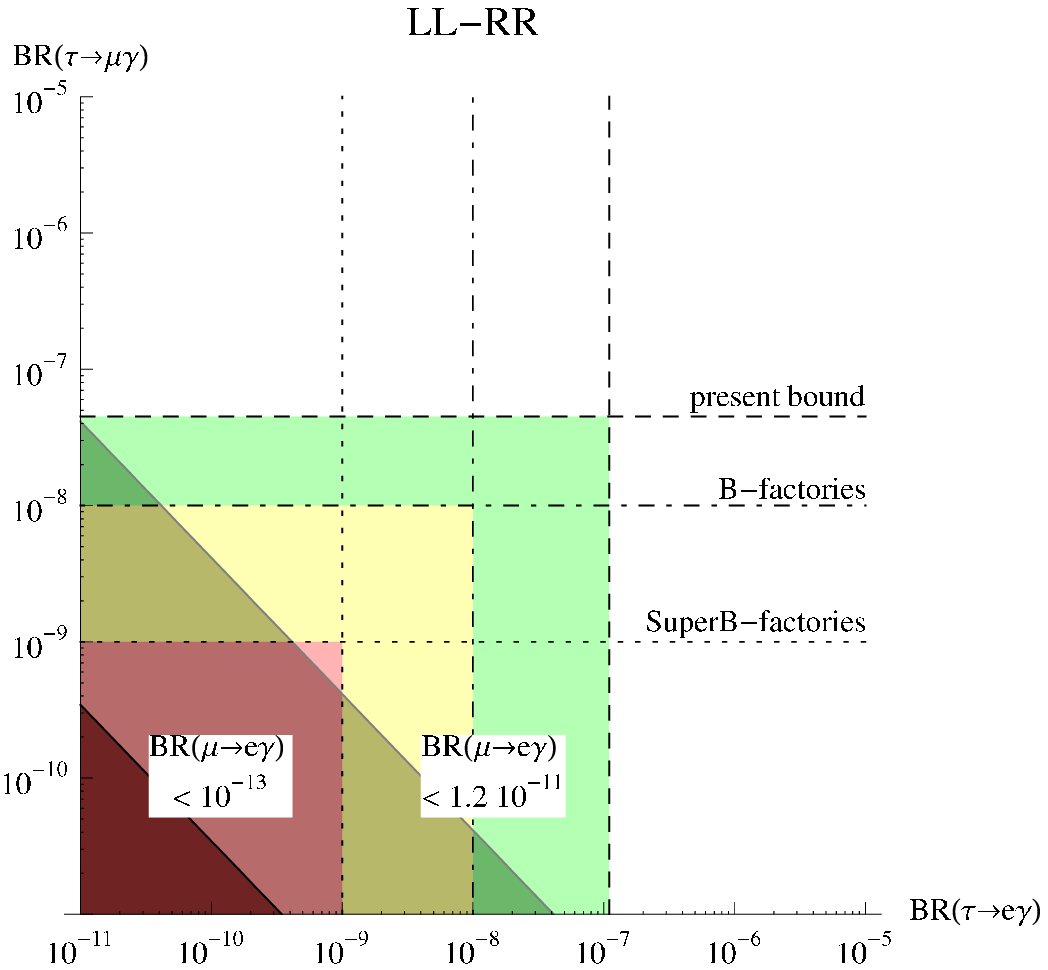,width=65mm} 
\epsfig{figure=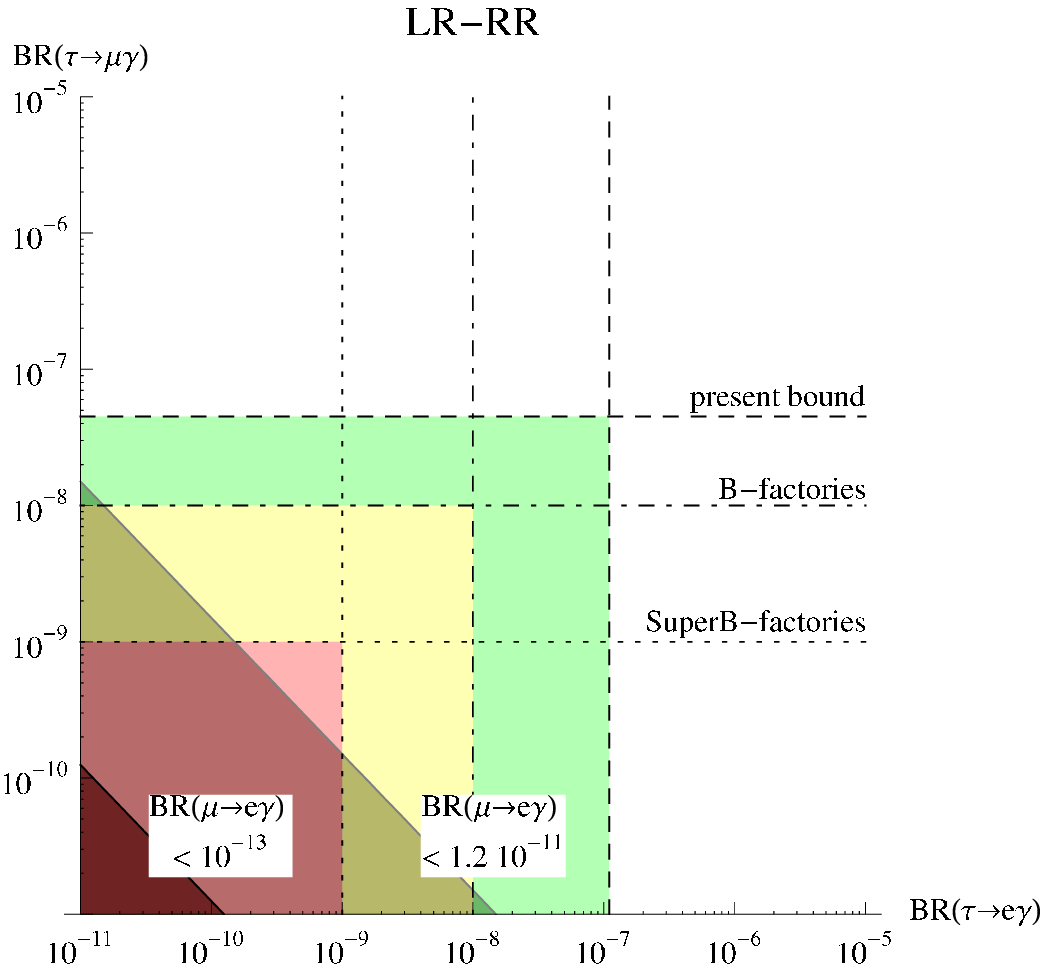,width=65mm}\\\\
\epsfig{figure=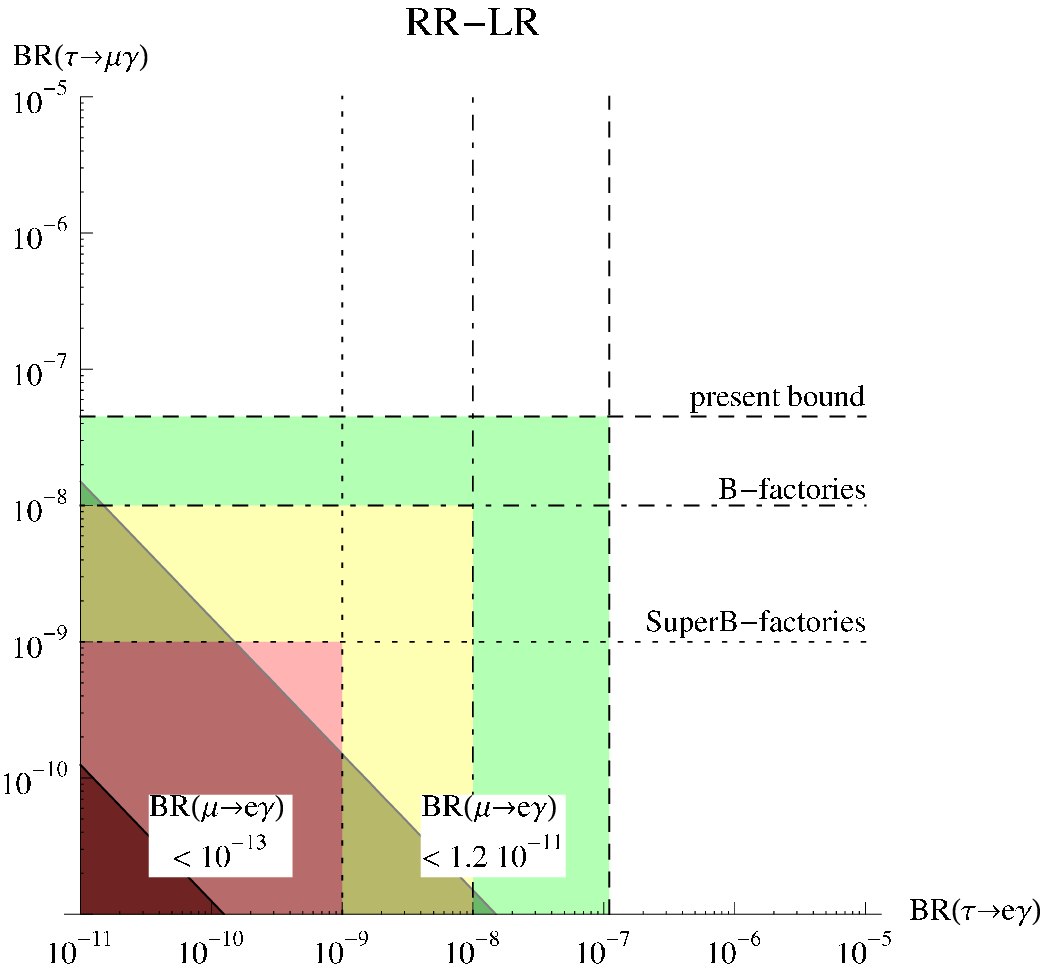,width=65mm} 
\epsfig{figure=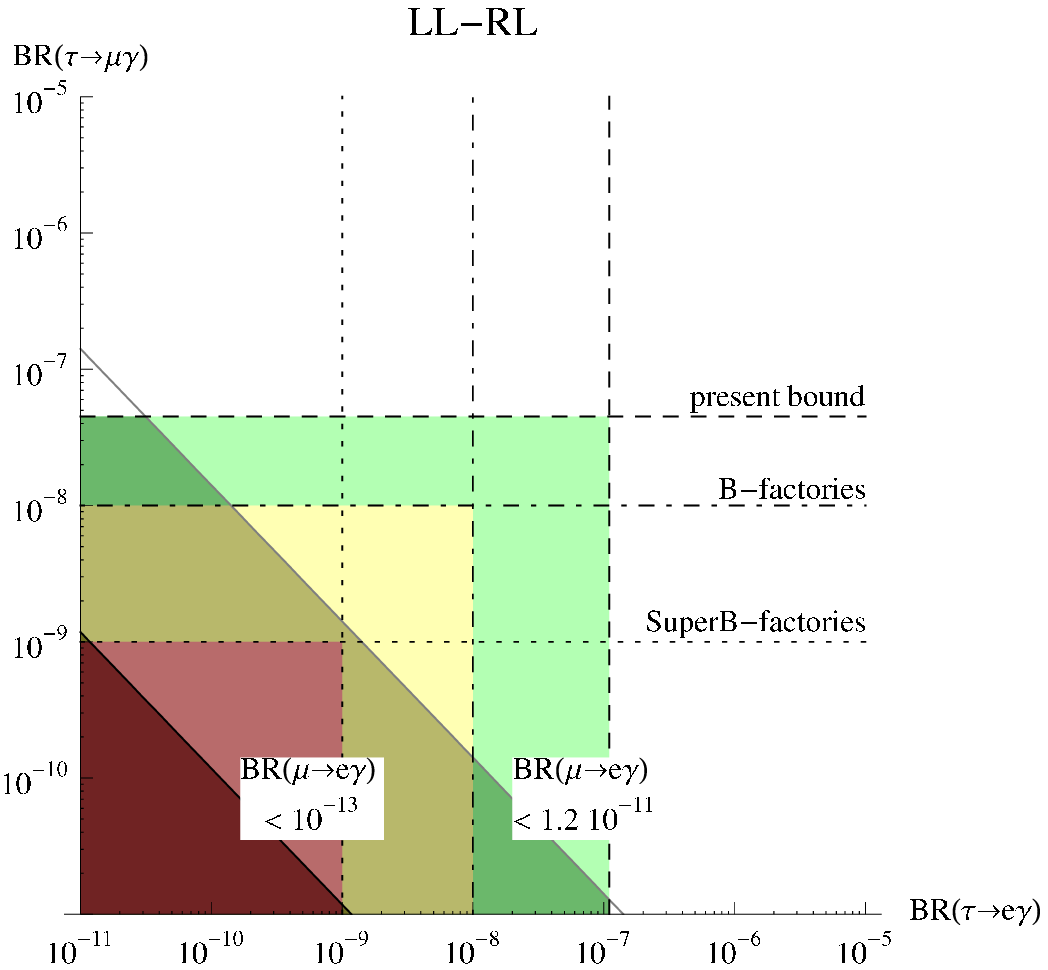,width=65mm}\\\\
\end{tabular}
\end{center}
\caption
{\small The same as Fig.\ref{fig:classI} but for Class II.
}
\label{fig:classII}
\end{figure}

\begin{figure}
\begin{center}
\begin{tabular}{c}
\epsfig{figure=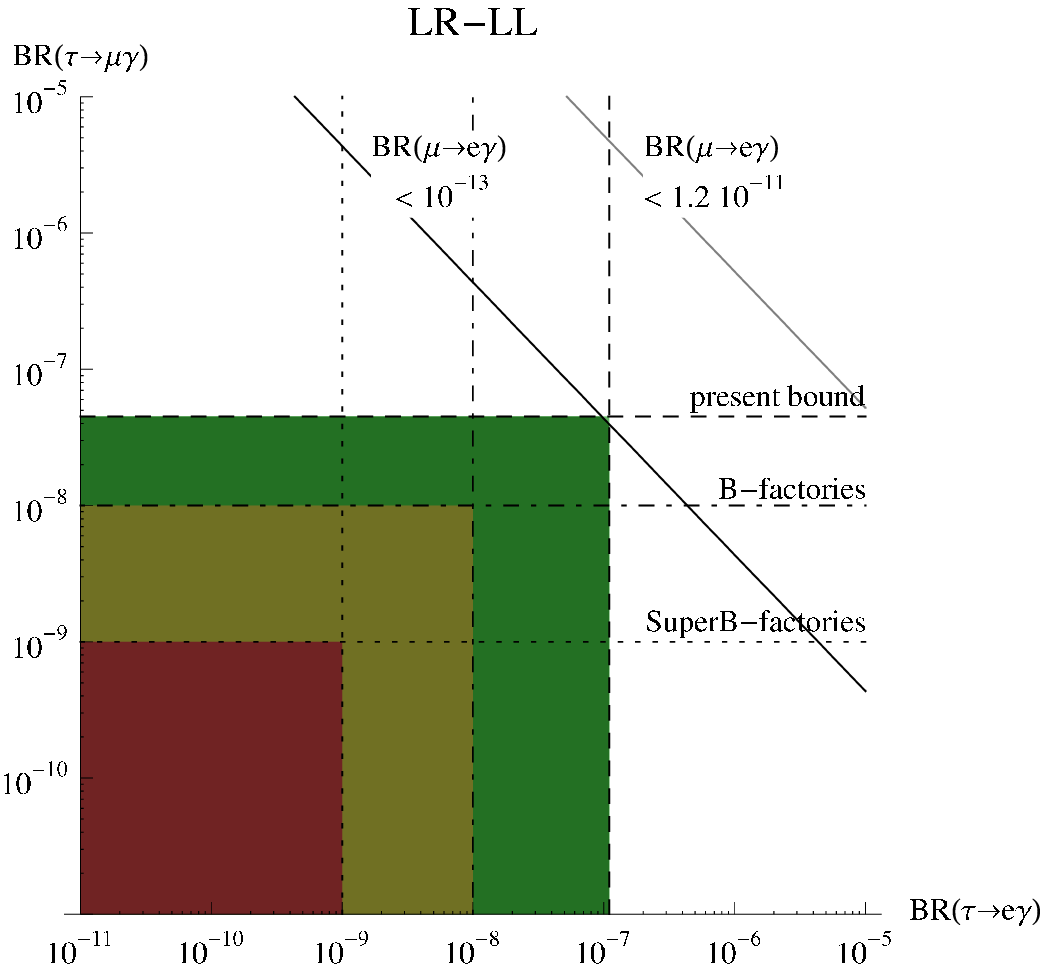,width=65mm} 
\epsfig{figure=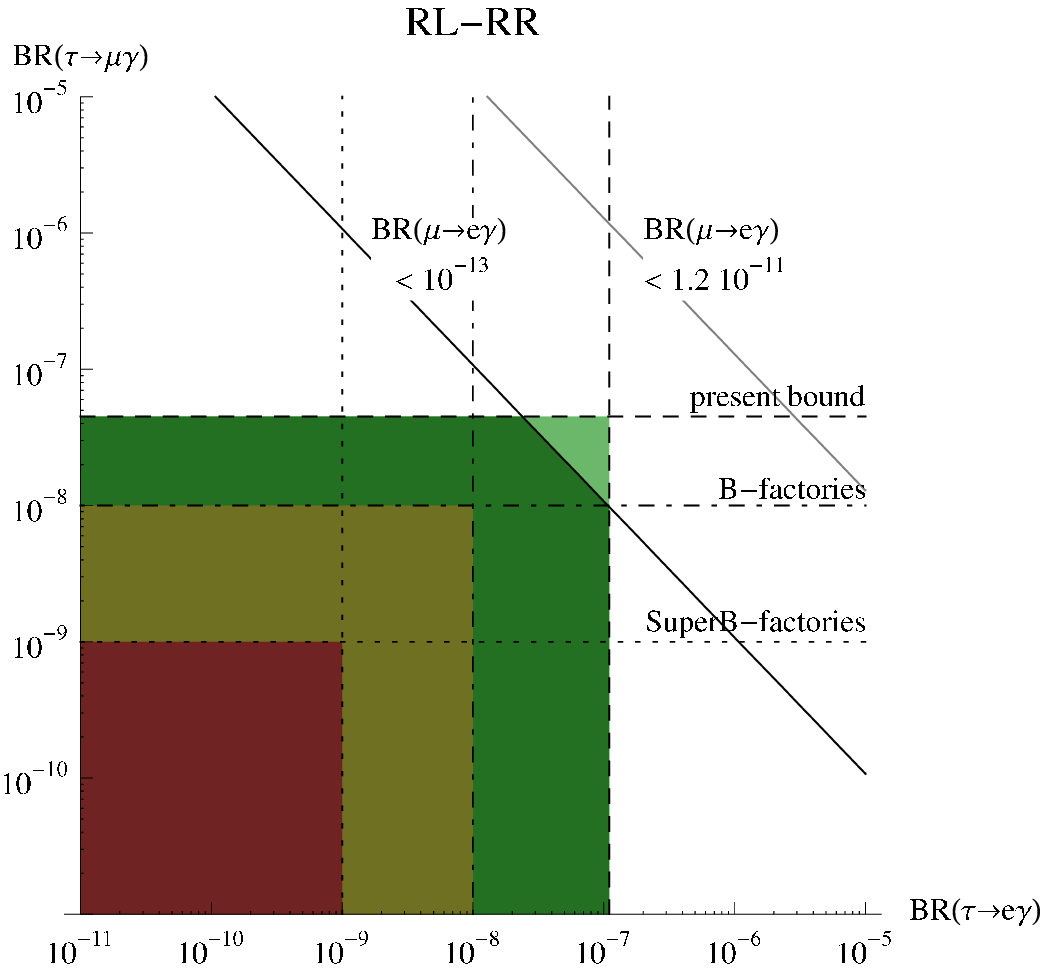,width=65mm}\\\\
\epsfig{figure=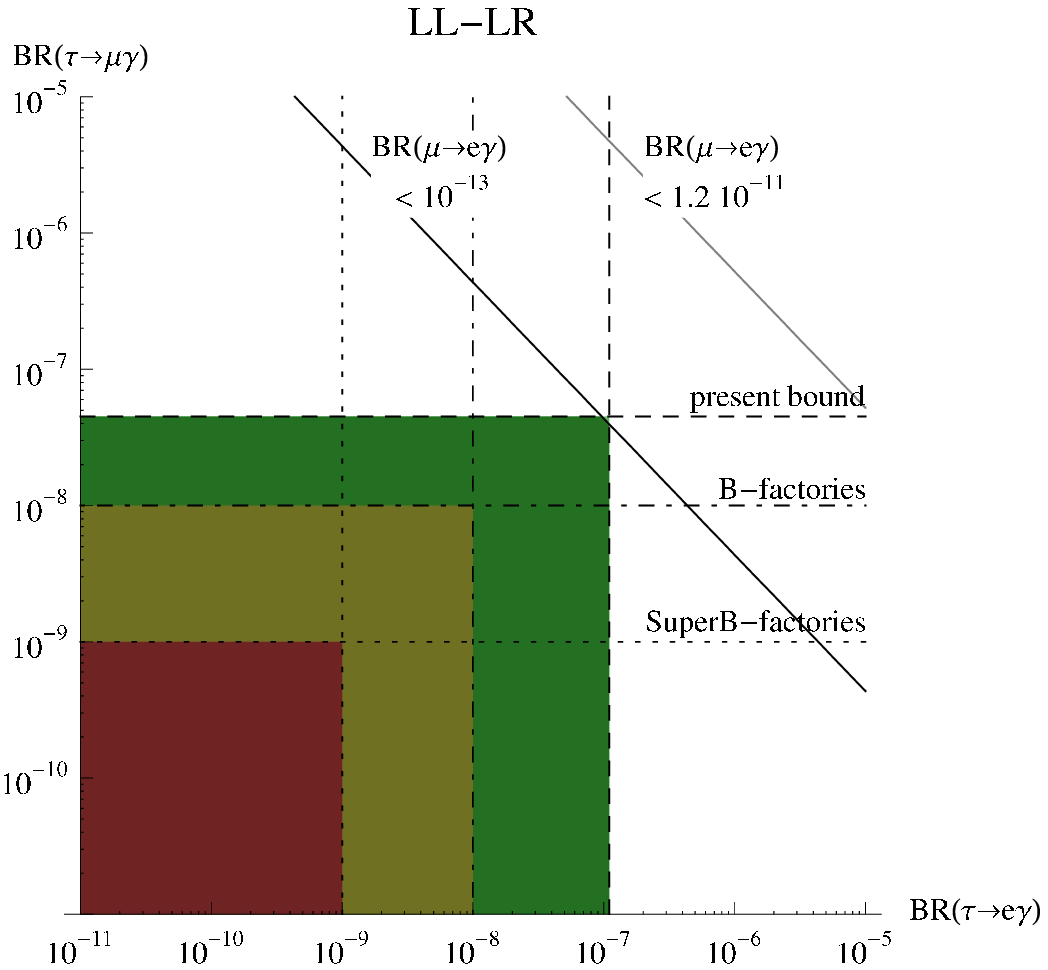,width=65mm} 
\epsfig{figure=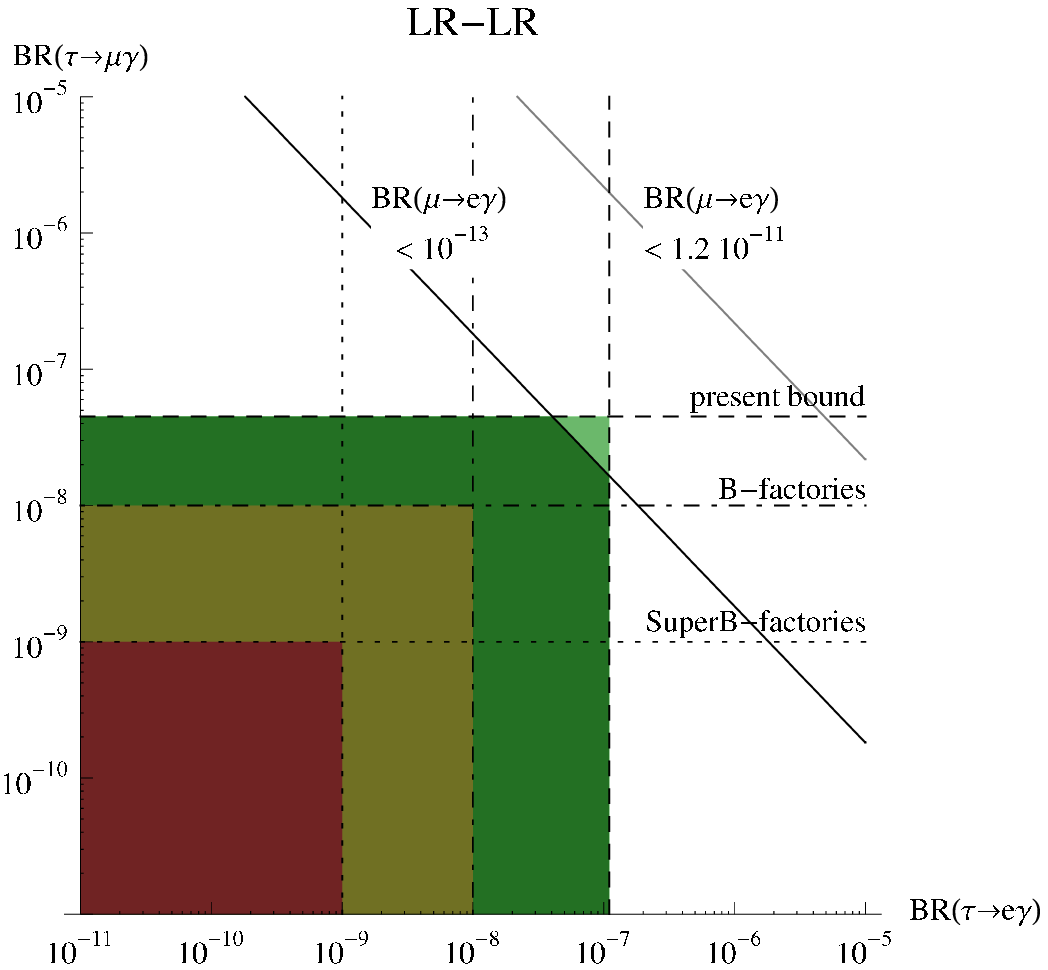,width=65mm}\\\\
\epsfig{figure=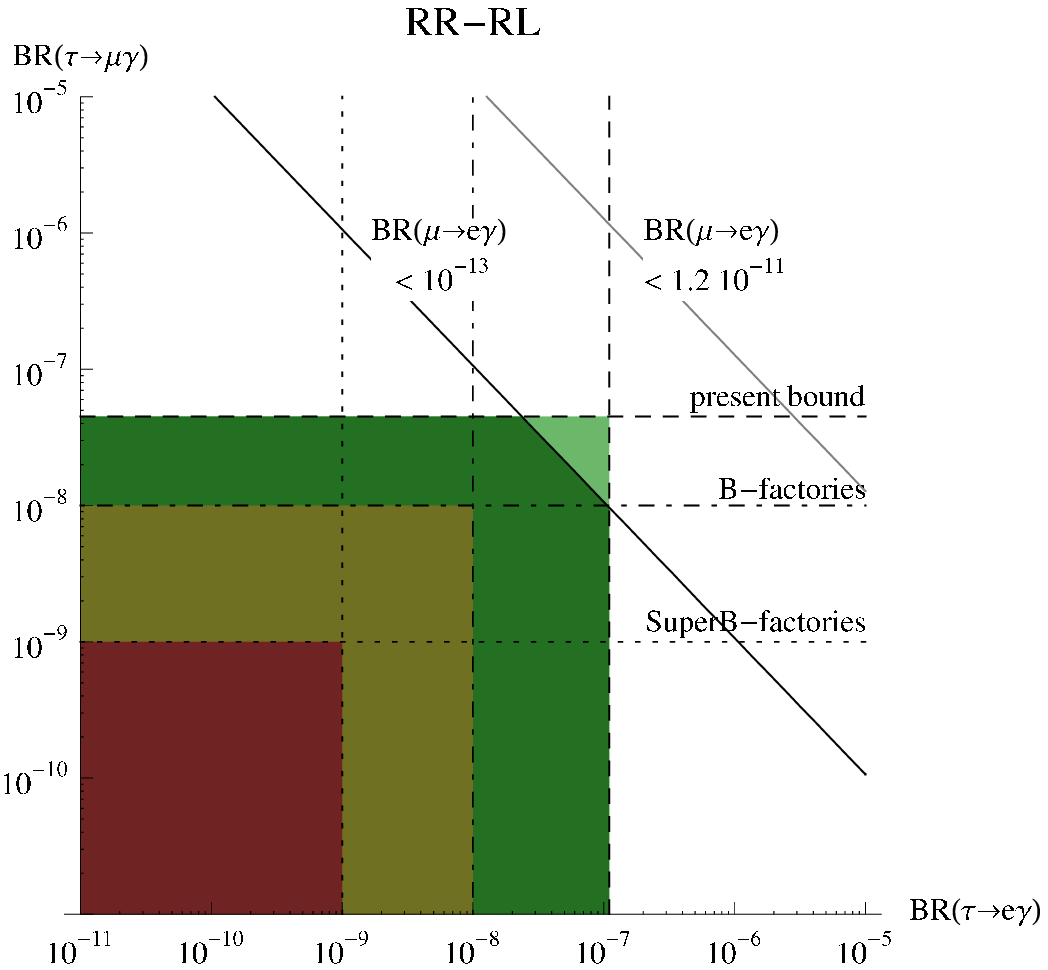,width=65mm} 
\epsfig{figure=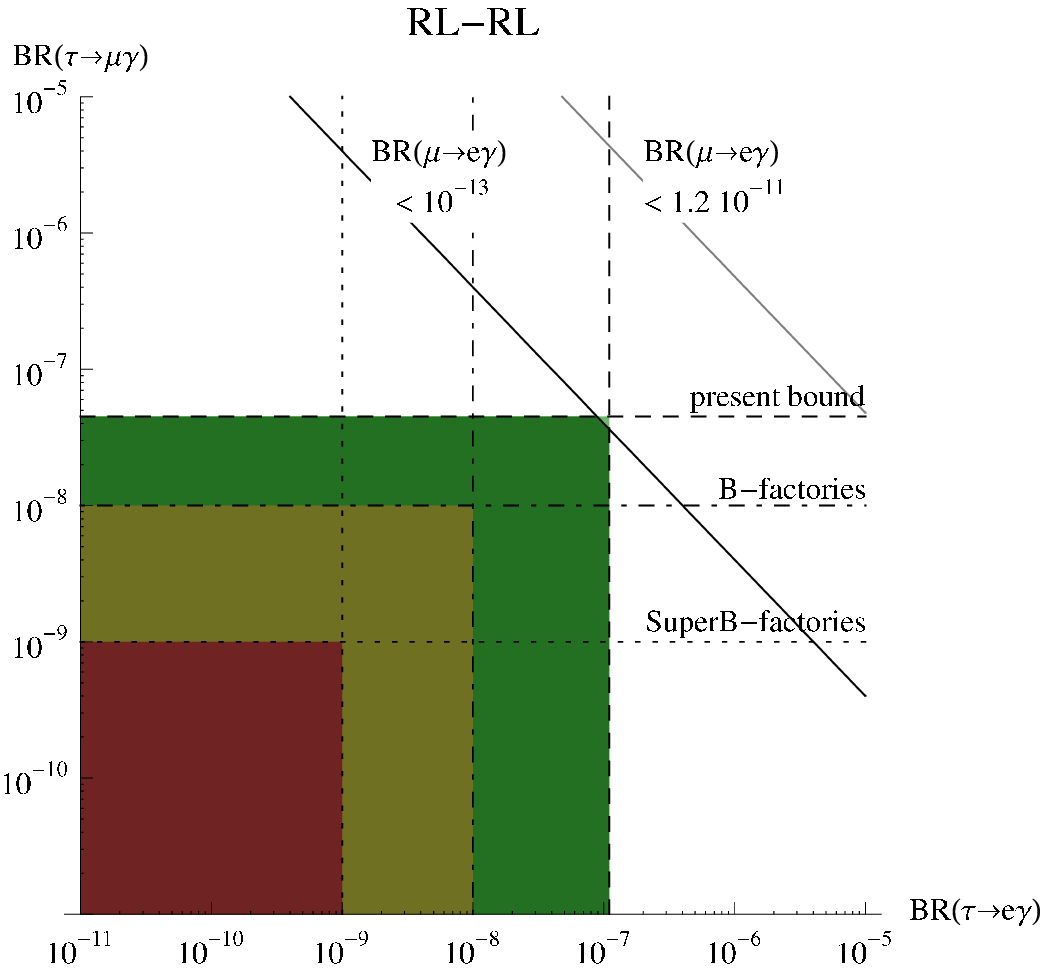,width=65mm}\\\\
\end{tabular}
\end{center}
\caption
{\small The same as Fig.\ref{fig:classI} but for Class III.
}
\label{fig:classIII}
\end{figure}

\begin{figure}
\begin{center}
\begin{tabular}{c}
\epsfig{figure=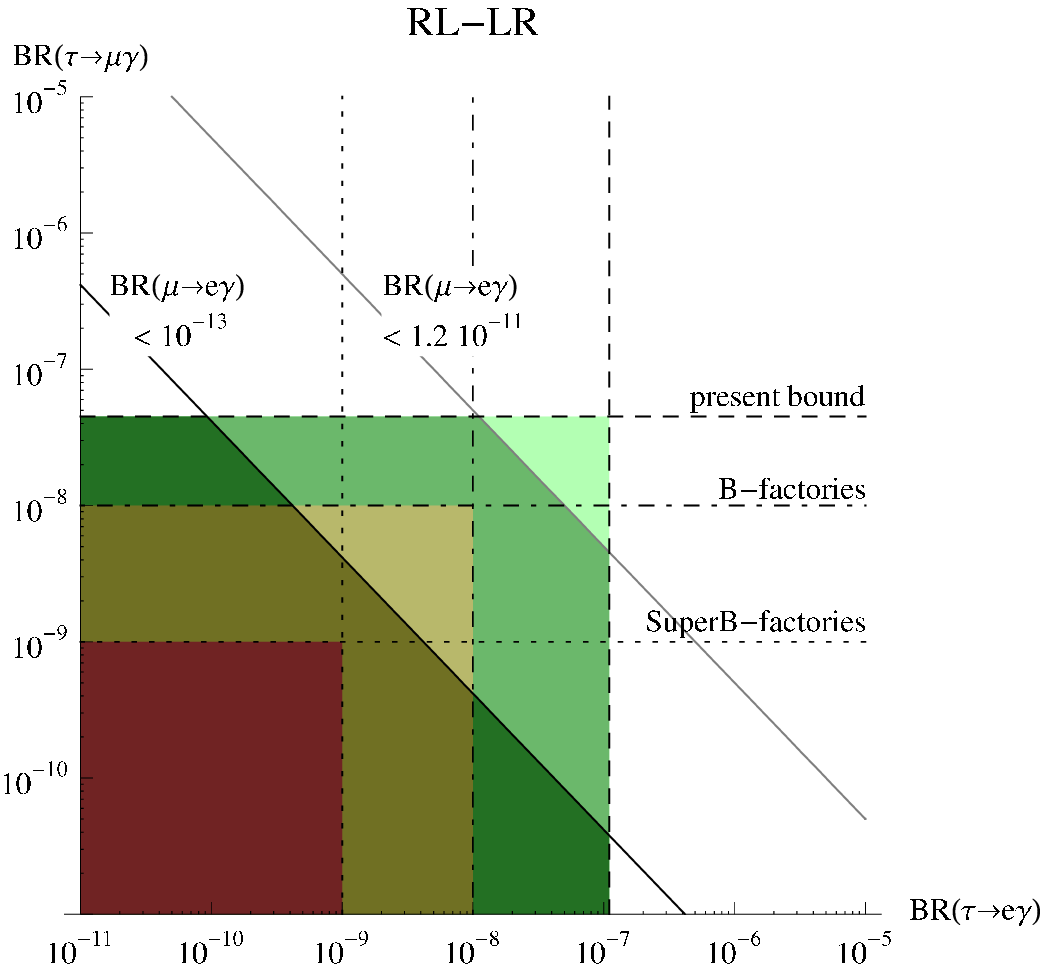,width=65mm} 
\epsfig{figure=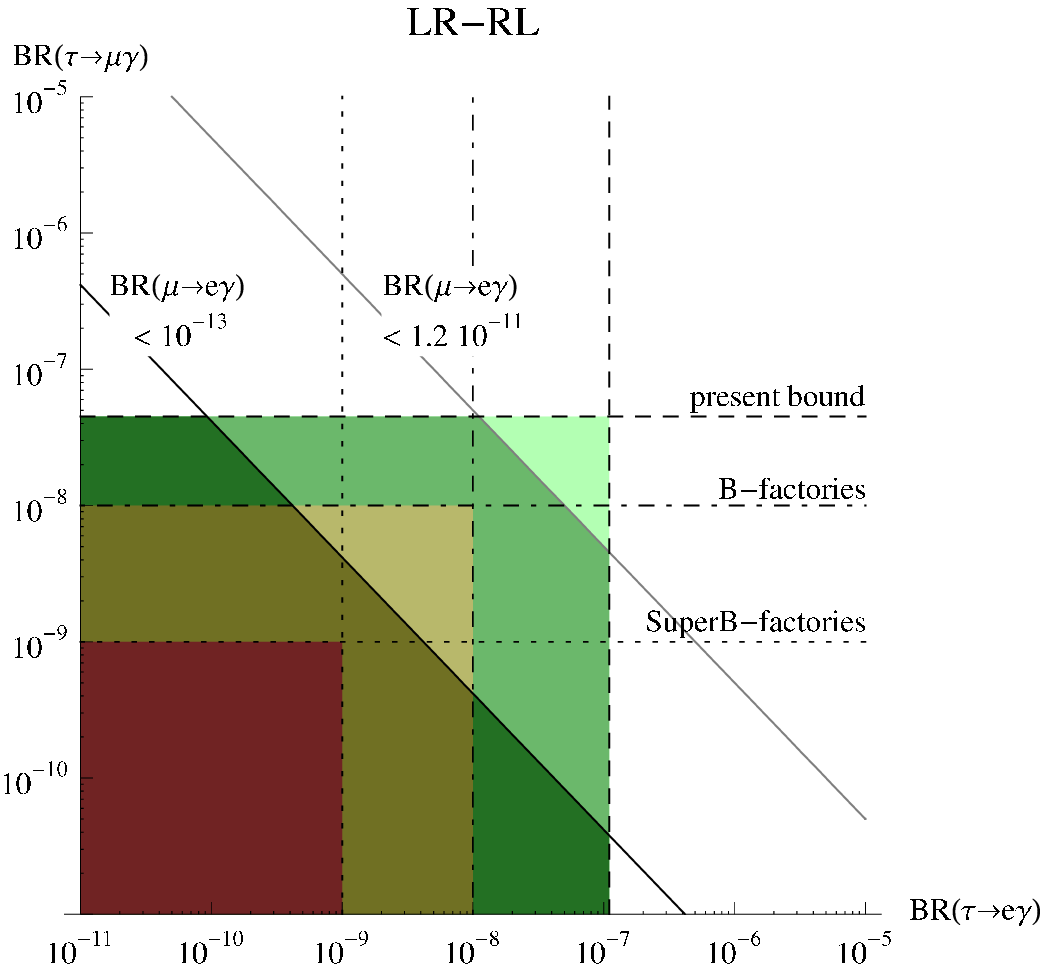,width=65mm}
\end{tabular}
\end{center}
\caption
{\small The same as Fig.\ref{fig:classI} but for Class IV.
}
\label{fig:classIV}
\end{figure}

In Figs.~\ref{fig:classI}-\ref{fig:classIV} 
we show, for Classes I-IV respectively, the allowed values for 
${\rm BR}(\tau\rightarrow e \gamma)$ and 
${\rm BR}(\tau\rightarrow \mu \gamma)$
in the MSSM for the mSUGRA benchmark point SPS1a; the
results for the SPS1b point are analogous and will not be
shown here. The area above (to the right of) the dashed line
at ${\rm BR}(\tau\rightarrow \mu \gamma)=4.5\times 10^{-8}$ 
(${\rm BR}(\tau\rightarrow e \gamma)=1.1\times 10^{-7}$) is 
excluded by the present experimental bounds on the rare tau decays.
On the other hand, the area above the diagonal line labeled 
${\rm BR}(\mu\rightarrow e\gamma)<1.2 \times 10^{-11}$ 
is excluded from the present experimental bound
on $\mu\rightarrow e\gamma$, as a consequence of
Eqs.~(\ref{class1}-\ref{class4}). 
The numerical results 
for these two benchmark points confirm our general expectations.
Namely, the theoretical constraints on the rare 
tau decays derived in this paper restrict values for
the branching ratios that are otherwise allowed by
present experiments, except for the models falling in
Class III.

The bounds Eqs.(\ref{class1}-\ref{class4}) 
also have implications for future searches for rare tau decays.
In Figs.~\ref{fig:classI}-\ref{fig:classIV} we show
with a dash-dotted line the projected sensitivity of
present $B$-factories to rare tau decays 
(${\rm BR}(\tau\rightarrow \mu \gamma),
{\rm BR}(\tau\rightarrow e \gamma)\gsim 10^{-8}$). Then, the area shaded
in green is the region of this parameter space accessible to
present $B$-factories. We find that
for Class II the region where {\it both} $\tau\rightarrow \mu \gamma$ 
and $\tau\rightarrow e \gamma$ could be discovered at present
$B$-factories is excluded. Therefore, if present $B$-factories
discovered both rare tau decays, only supersymmetric models
falling in Classes III, IV and Class I (for the case with LL-LL 
mass insertions)
would be allowed. This conclusion will be strengthened 
if the MEG experiment
at PSI reaches the projected sensitivity ${\rm BR}(\mu\rightarrow e \gamma)
\sim 10^{-13}$ without finding a positive signal.
If this is the case, the observation
of both tau rare decays at present $B$-factories would point
to an origin of the tau flavour violation falling only in Class III.
The same rationale could be applied to the future searches of rare tau
decays at the projected super$B$-factories.
In Figs. ~\ref{fig:classI}-\ref{fig:classIV} we also show
as a yellow shaded area 
the region of the parameter space accessible to the projected 
super$B$-factories (${\rm BR}(\tau\rightarrow \mu \gamma),
{\rm BR}(\tau\rightarrow e \gamma)\gsim 10^{-9}$). Whereas the present
bound on $\mu\rightarrow e\gamma$ only has implications for
the projected super$B$-factories for the models falling in Class II,
if the bound on $\mu\rightarrow e\gamma$ is improved to the level
of $10^{-13}$ our results will also be relevant for the models
falling in Classes I and IV.

It is interesting to note that two of the most widely studied
scenarios generating sizable rates for the rare decays,
namely the supersymmetric see-saw model and the minimal SU(5)
grand unified model, fall in Class I. To be precise,
the supersymmetric see-saw model generates flavour violation
in the LL sector~\cite{Borzumati:1986qx} and the minimal SU(5) model,
 in the RR sector~\cite{Barbieri:1994pv}.
For Class I the bound Eq.(\ref{class1}) is quite stringent
and disfavours the possibility of observing both rare tau
decays at present $B$-factories for a generic point of the
mSUGRA parameter space. Furthermore, the bounds derived in this
paper for the MSSM are conservative and typically 
become more stringent as the physics that generates the
flavour violation is specified. Indeed, as was shown
in \cite{Ibarra:2008uv}, two loop effects induced by right-handed 
neutrinos in the supersymmetric see-saw model generate
the off-diagonal terms $(\mLs)_{12}$, $(\Ae)_{12}$ and $(\Ae)_{21}$,
which contribute through a single mass insertion to
${\rm BR}(\mu\rightarrow e\gamma)$ in addition to the double
mass insertion contribution considered in the present work.

To finish this section, let us review 
other theoretical constraints on the rare tau decays
that have been derived in the literature for the MSSM. Interesting
bounds on the branching ratios of the rare lepton
decays were derived 
in \cite{Casas:1996de} from requiring absence of charge breaking
minima or unbounded from below directions in the 
effective potential. The respective resulting bounds on the LR and RL
mass insertions read, 
\begin{eqnarray}
|\Delta^{\rm (LR)}_{ij}|, |\Delta^{\rm (RL)}_{ji}| &\leq& m_k
\left[(\mes)_{ii}+(\mLs)_{jj}+m_{H_d}^2\right]^{1/2} \;, \nonumber \\
|\Delta^{\rm (LR)}_{ij}|, |\Delta^{\rm (RL)}_{ji}|  &\leq& m_k 
\left[(\mes)_{ii}+(\mLs)_{jj}+(\mLs)_{nn}\right]^{1/2},
~~~~~n\neq i,j \;,
\end{eqnarray}
where $m_k$ is the lepton mass, $k={\rm Max}\ (i,j)$,
and $m_{H_d}^2$ is the down-type Higgs mass squared.
Substituting these bounds on the mass insertions in Eq.(\ref{1MI}) 
one finally obtains the
following approximate constraint on the radiative tau decays:
\begin{equation}
{\rm BR}(\tau\rightarrow \l \gamma)
\lsim \frac{3 \alpha^3}{G_F^2 \widetilde m^4}
\;{\rm BR}(\tau \rightarrow \l \nu_\tau \bar \nu_{\l})
\sim 6\times 10^{-8} \left(\frac{\widetilde m}{400\,{\rm GeV}}\right)^{-4}\;,
\label{CCB}
\end{equation}
which is comparable, for $\widetilde m \sim 400$ GeV, to the present 
experimental bounds on the rare tau decays. In particular, 
we obtain for the SPS1a (SPS1b) benchmark point,
\mbox{${\rm BR}(\tau\rightarrow \l \gamma)\lsim 9\times 10^{-7}$ $(10^{-7})$ }
in the case of the LR mass insertion and 
${\rm BR}(\tau\rightarrow \l \gamma)\lsim 9\times 10^{-7}$ $(2\times 10^{-7})$
for the case of the RL mass insertion.

If the origin of the lepton flavour violation in the rare tau
decays could be pinpointed to the LR or the RL sector, the 
observation of $\tau\rightarrow \mu \gamma$ or $\tau\rightarrow e \gamma$
in the near future would set, following Eq.(\ref{CCB}), 
an {\it upper bound} on the scalar masses,
$\widetilde m\lsim 400$ GeV, in order to avoid 
the appearance of dangerous charge breaking minima or unbounded 
from below directions in the effective potential. Remarkably,
the constraints on the rare tau decays derived in this paper 
could help to pinpoint the origin of the lepton 
flavour violation. As was argued before, if both 
$\tau\rightarrow e\gamma$ and  $\tau\rightarrow \mu \gamma$ 
were observed in the near future, models falling in Class III, and
possibly also in Class IV, would be favoured over models falling
in Classes I and II, especially if the experimental bound
on ${\rm BR}(\mu\rightarrow e\gamma)$ reaches the level of $10^{-13}$. 
Therefore, since models falling in Classes III 
or IV always involve a LR and/or a RL mass insertion, the bound 
Eq.(\ref{CCB}) would apply at least for one of the rare decays,
and accordingly an upper bound on the scalar masses would
follow. Namely, if future experiments show that 
${\rm BR}(\mu\rightarrow e\gamma)\leq 10^{-13}$ but
${\rm BR}(\tau\rightarrow \l \gamma)> 10^{-8}$, it would follow
that $\widetilde m\lsim 700$ GeV 
from requiring absence of charge breaking
minima or unbounded from below directions in the 
effective potential. 

\section{The effective field theory approach}

In this section we will derive, pursuing an
effective field theory approach, a conservative bound on
${\rm BR}(\mu\rightarrow e\gamma)$ in terms of the branching
ratios for the radiative tau decays. The resulting bound
will be therefore completely model independent.

\begin{figure}
 \centering \small
 \begin{minipage}{30mm}
  \epsfig{figure=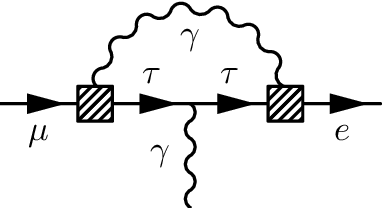,width=30mm}
  \centering (A1)
 \end{minipage}\hspace{3mm}
 \begin{minipage}{30mm}
  \epsfig{figure=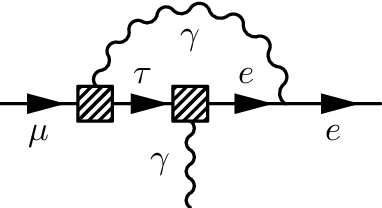,width=30mm}
  \centering (A2)
 \end{minipage}\hspace{3mm}
 \begin{minipage}{30mm}
  \epsfig{figure=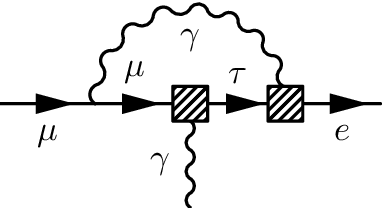,width=30mm}
  \centering (A3)
 \end{minipage}
 \\ \vspace{.4cm}
 \begin{minipage}{30mm}
  \epsfig{figure=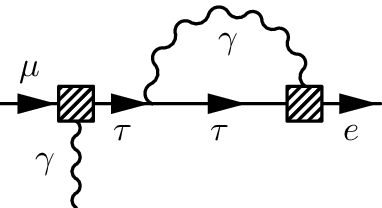,width=30mm}
  \centering (B1)
 \end{minipage}\hspace{3mm}
 \begin{minipage}{30mm}
  \epsfig{figure=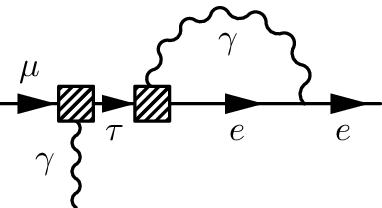,width=30mm}
  \centering (B2)
 \end{minipage}\hspace{3mm}
 \begin{minipage}{30mm}
  \epsfig{figure=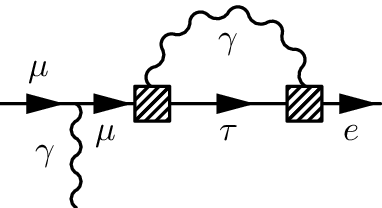,width=30mm}
  \centering (B3)
 \end{minipage}
\\ \vspace{.4cm}
 \begin{minipage}{30mm}
  \epsfig{figure=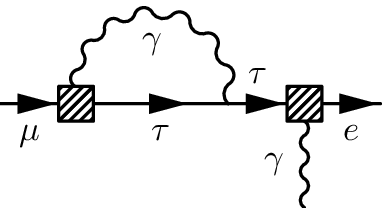,width=30mm}
  \centering (C1)
 \end{minipage}\hspace{3mm}
 \begin{minipage}{30mm}
  \epsfig{figure=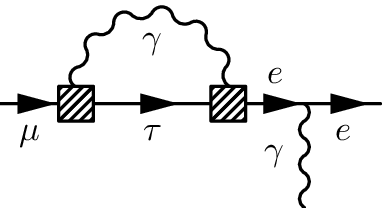,width=30mm}
  \centering (C2)
 \end{minipage}\hspace{3mm}
 \begin{minipage}{30mm}
  \epsfig{figure=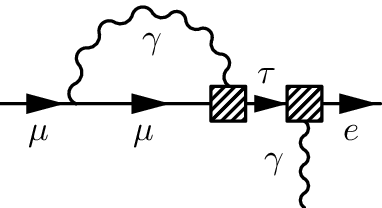,width=30mm}
  \centering (C3)
 \end{minipage}
 \caption{\small One loop Feynman diagrams that induce the process 
          $\mu \rightarrow e \gamma$ from the effective operators
          that induce $\tau \rightarrow \mu \gamma$
          and $\tau \rightarrow e \gamma$.}
 \label{fig:EffOpermeg}
\end{figure}

Our starting point is the electromagnetic transition amplitude
for the processes $\tau \rightarrow \mu \gamma^*$ and $\tau \rightarrow e \gamma^*$,
Eq.~(\ref{transition}). If both transitions exist in Nature,
the transition $\mu \rightarrow e \gamma^*$ will be automatically
induced through the nine diagrams shown in Fig.~\ref{fig:EffOpermeg}.
Among these,
the diagrams (B3) and (C2) do not contribute to the dipole form 
factors $f_{M1}^{\mu e}$, $f_{E1}^{\mu e}$, which are
the only ones that induce the process $\mu \rightarrow e\gamma$.
On the other hand,
since the photon circulating in the loop is off-shell, 
all the form factors that induce
the electromagnetic tau transition (monopole and dipole) will
contribute to $f_{M1}^{\mu e}$, $f_{E1}^{\mu e}$.
However, in order to derive a bound of the form 
${\rm BR}(\mu \rightarrow e \gamma)\gsim 
C\times {\rm BR}(\tau \rightarrow \mu \gamma)
{\rm BR}(\tau \rightarrow e \gamma)$,
we will be interested just in the contribution from the dipole operators,
which are the only ones that induce the processes 
$\tau\rightarrow \mu\gamma$ and $\tau\rightarrow e\gamma$. 
We estimate that the dipole form factors satisfy the following relations:
\begin{eqnarray}
\left|f_{M1}^{\mu e}\right|&\gsim& \frac{9\alpha}{2 \pi} 
\frac{m^3_\tau}{m_\mu} \left|f_{E1}^{\tau e*} f_{E1}^{\tau \mu}- 
f_{M1}^{\tau e*} f_{M1}^{\tau \mu}\right|\log \frac{\Lambda}{m_\mu}\;,\nonumber  \\
\left|f_{E1}^{\mu e}\right|&\gsim& \frac{9\alpha}{2 \pi} 
\frac{m^3_\tau}{m_\mu} \left|f_{E1}^{\tau e*} f_{M1}^{\tau \mu}- 
f_{M1}^{\tau e*} f_{E1}^{\tau \mu}\right|\log \frac{\Lambda}{m_\mu}\;,
\end{eqnarray}
from where it follows that
\begin{eqnarray}
{\rm BR}(\mu\rightarrow e\gamma) &\gsim&
\frac{1944 \pi \alpha^3}{G_F^2}
\frac{m^6_\tau}{m^2_\mu}
\left[
(|f_{E1}^{\tau \mu}|^2+|f_{M1}^{\tau \mu}|^2)
(|f_{E1}^{\tau e}|^2+|f_{M1}^{\tau e}|^2)\right. \nonumber \\
&-&\left.4 {\rm Re}(f_{E1}^{\tau e} f_{M1}^{\tau e *})
{\rm Re}(f_{E1}^{\tau \mu} f_{M1}^{\tau \mu *})
\right]
\log^2 \frac{\Lambda}{m_\mu}\;,
\end{eqnarray}
being $\Lambda$ a cutoff.
Assuming that each rare tau decay is dominated by just one
of the dipole form factors, either the electric or the magnetic,
one finally obtains
\begin{eqnarray}
{\rm BR}(\mu\rightarrow e\gamma) &\gsim&
\frac{27\, G_F^2\, \alpha}{128\, \pi^5}\frac{m^6_\tau}{m^2_\mu}
\log^2 \frac{\Lambda}{m_\mu}
\frac{{\rm BR}(\tau\rightarrow \mu\gamma)}
{{\rm BR}(\tau \rightarrow \mu \nu_\tau \bar \nu_\mu)}
\frac{{\rm BR}(\tau\rightarrow e\gamma)}
{{\rm BR}(\tau \rightarrow e \nu_\tau \bar \nu_e)}\nonumber \\
&\sim&4\times 10^{-23}
\left(\frac{{\rm BR}(\tau\rightarrow \mu\gamma)}{4.5\times10^{-8}}\right)
\left(\frac{{\rm BR}(\tau\rightarrow e\gamma)}{1.1\times10^{-7}}\right)\;,
\end{eqnarray}
where we have used $\Lambda=1$ TeV.
The resulting bound is too weak to have any practical
application, although it is has the theoretically interest
of setting an {\it absolute} lower bound on ${\rm BR}(\mu\rightarrow e\gamma)$
in terms of the rare tau decays. As the fundamental
theory that generates the effective dipole operators
becomes specified, new contributions to the rare
muon decay will typically arise, thus strengthening
considerably the previous bound. This is the case in particular
for the Minimal Supersymmetric Standard Model 
considered in the previous section: the one loop
diagrams that induce the process $\mu\rightarrow e\gamma$
in the effective theory approach, Figs.~\ref{fig:EffOpermeg},
correspond to much more suppressed three loop diagrams once
the complete theory has been specified.

\begin{figure}
\centering
\begin{minipage}{30mm}
\epsfig{figure=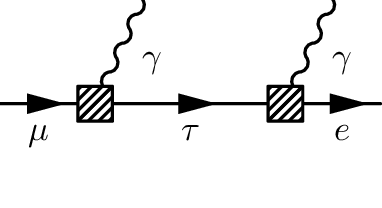,width=30mm}
\end{minipage}
\caption
{\small Feynman diagram that induces the process 
$\mu \rightarrow e \gamma\gamma$ from the effective operators
that induce $\tau \rightarrow \mu \gamma$
and $\tau \rightarrow e \gamma$.
}
\label{fig:EffOperme2g}
\end{figure}

Using the same effective theory approach it is possible
to compute also a lower bound on the branching ratio for the
process $\mu \rightarrow e\gamma\gamma$, which is induced
by the diagram shown in Fig.~\ref{fig:EffOperme2g}.
The result is
\begin{equation}
{\rm BR}(\mu \rightarrow e\gamma\gamma) \gsim
\frac{8 m^2_\mu m^2_\tau \pi^2 \alpha^2}{5 G^2_F}
\left[
(|f_{E1}^{\tau \mu}|^2+|f_{M1}^{\tau \mu}|^2)
(|f_{E1}^{\tau e}|^2+|f_{M1}^{\tau e}|^2)
-4 {\rm Re}(f_{E1}^{\tau e} f_{M1}^{\tau e *})
{\rm Re}(f_{E1}^{\tau \mu} f_{M1}^{\tau \mu *})
\right]\,.
\end{equation}
As before, this bound can be rewritten in terms of the
branching ratios of the radiative tau decays, yielding
\begin{eqnarray}
{\rm BR}(\mu\rightarrow e\gamma\gamma) &\gsim&
\frac{G_F^2 m^2_\mu m^2_\tau}{5760 \pi^4}  
\frac{{\rm BR}(\tau\rightarrow \mu\gamma)}
{{\rm BR}(\tau \rightarrow \mu \nu_\tau \bar \nu_\mu)}
\frac{{\rm BR}(\tau\rightarrow e\gamma)}
{{\rm BR}(\tau \rightarrow e \nu_\tau \bar \nu_e)}  \nonumber \\
&\sim&10^{-30}
\left(\frac{{\rm BR}(\tau\rightarrow \mu\gamma)}{4.5\times10^{-8}}\right)
\left(\frac{{\rm BR}(\tau\rightarrow e\gamma)}{1.1\times10^{-7}}\right)\;,
\end{eqnarray}
again far below the experimental upper bound, 
${\rm BR}(\mu\rightarrow e\gamma\gamma) <7.2\times 10^{-11}$
~\cite{Grosnick:1986pr}.

\section{Conclusions}

We have derived in this paper theoretical constraints 
on the branching ratios of the rare tau decays
of the form ${\rm BR}(\mu \rightarrow e \gamma)\gsim 
C\times {\rm BR}(\tau \rightarrow \mu \gamma)
{\rm BR}(\tau \rightarrow e \gamma)$ in the 
Minimal Supersymmetric Standard Model and in
a completely general setup, pursuing an effective
field theory approach. 

We have argued that in the MSSM the 
observation of both rare tau decays implies, barring 
cancellations, a non-vanishing rate for the
process $\mu \rightarrow e \gamma$ through the double
mass insertion in the slepton propagator. We have 
cataloged the sixteen possibilities for the double
mass insertion in four classes, according to their dependence
on $\tan\beta$, the fermion masses and the overall size
of the scalar masses, which are the three parameters to
which the constant $C$ is most sensitive to, and we have shown 
that for a wide class of models our bound constrains values
for the branching ratios of the rare tau decays
that are otherwise allowed by present experiments. 
We have shown that if present $B$-factories
observe both $\tau \rightarrow \mu \gamma$ and
$\tau \rightarrow e \gamma$, the underlying possible
sources of flavour violation would be restricted
to our Class III, and  possibly Class IV, 
unless the supersymmetric parameters take special values. 
This conclusion would be strengthened if the MEG experiment at PSI 
reaches the sensitivity of $10^{-13}$ for ${\rm BR}(\mu\rightarrow e\gamma)$ 
without finding a positive signal. We have also discussed the 
complementarity of the constraints on the rare tau decays derived 
in this paper and the constraints stemming from requiring absence
of charge breaking minima or unbounded from below directions
in the effective potential.

Finally, we have derived for completeness 
theoretical constraints on the rare tau
decay following an effective theory approach.
The resulting bounds are too weak to have any
practical interest, although they have the
theoretical interest of setting absolute
bounds on the rates of $\mu \rightarrow e \gamma$ and 
$\mu \rightarrow e \gamma\gamma$ in terms of the rates of 
$\tau \rightarrow \mu \gamma$ and $\tau \rightarrow e \gamma$.

\section*{Acknowledgements}
We are grateful to Sacha Davidson, Paride Paradisi and 
especially to Jos\'e Ram\'on Espinosa for interesting 
discussions and suggestions.
This research was supported by the DFG cluster of
excellence Origin and Structure of the Universe and by the 
SFB-Transregio 27 ``Neutrinos and Beyond''.

\end{document}